\documentclass{aa}  

\usepackage{graphicx}
\usepackage{txfonts}
\usepackage{lipsum}
\usepackage{subcaption}         
\usepackage{xcolor}             
\usepackage{lscape}             
\usepackage{placeins}           
                                
\begin{document}

   \title{Cosmogenic neutrinos from FRII galaxies as potential origin of the ultra-high-energy KM3-230213A event}


%
%
%

   \author{D. Allard\inst{1}\corrauth{allard@apc.in2p3.fr}
        \and B. Baret\inst{1}\email{baret@apc.in2p3.fr}
        \and N. Globus\inst{2}\email{globus@astro.unam.mx}
        \and E. Parizot\inst{1}\email{parizot@apc.in2p3.fr}
        }

   \institute{Université Paris Cité, CNRS, Laboratoire Astroparticule et Cosmologie, F-75013 Paris, France 
   \and Instituto de Astronomía, Universidad Nacional Autónoma de México, km 107 Carretera Tijuana-Ensenada, 22860, Ensenada, México}

   \date{Received September 30, 20XX}

 
  \abstract
   {}
  {We investigate whether the ultra-high-energy neutrino KM3-230213A can be interpreted as a cosmogenic neutrino produced by ultra-high-energy cosmic rays (UHECRs) accelerated in the lobes of FRII radio galaxies.}
  {We model the UHECR, cosmogenic neutrino and photon fluxes expected at Earth using a recent luminosity-dependent density evolution of radio galaxies, empirical relations between radio luminosity and jet kinetic power, and standard assumptions for the UHECR output of FRII lobes. The FRII contribution to the UHECR population is derived self-consistently from the observed luminosity function, rather than imposed as a fixed normalization. The propagation of UHECRs and the production of secondary particles are computed with well-established numerical tools.}
  {The predicted cosmogenic neutrino flux is compatible with that inferred from the detection of KM3-230213A, while remaining consistent with current UHECR and gamma-ray constraints. According to our models, the full GRAND observatory ($200\,000~\rm km^2$) should detect between $\sim50$ and $\sim135$ neutrinos above $10^{17}$~eV in ten years, allowing the diffuse UHE neutrino spectrum to be characterized. In contrast, the detection of individual FRII sources or statistically significant correlations with FRII catalogs is likely to remain challenging.
  At the highest energies, UHECR composition and anisotropy measurements, in particular those related to the nearby radio galaxy Cygnus~A, should provide complementary tests of this scenario. More generally, progress will likely rely on the combination of multimessenger observations with improved astrophysical constraints on particle acceleration and jet composition in FRII radio galaxies.}
   {}

   \keywords{Astroparticle physics --
                Galaxies: luminosity function, mass function --
                Neutrinos
               }

   \maketitle

\nolinenumbers

\section{Motivations}

Cosmogenic neutrinos \citep{Berezinsky1969} are produced during the intergalactic propagation of ultra-high-energy cosmic rays (UHECRs) through their interactions with photon backgrounds. Their production is therefore unavoidable, making them one of the cornerstones of multimessenger studies of UHECRs. The expected cosmogenic neutrino flux depends strongly on both the cosmological evolution of UHECR sources (see, e.g., \citet{Seckel2005}) and the maximum energy to which they accelerate cosmic rays (see, e.g., \citet{Engel2001}). In standard scenarios, most cosmogenic neutrinos above $10^{17}$~eV are produced through pion production on photons of the cosmic microwave background (CMB), a process that requires protons (or nuclei) with energies of at least a few $10^{19}$~eV per nucleon. The mass-composition measurements reported by the Pierre Auger Observatory \citep{Abraham2010, Aab2014, Aab2014b} therefore appeared to reduce the prospects for observing large cosmogenic neutrino fluxes. The gradual shift toward a heavier composition above the ankle is generally interpreted as evidence that the maximum rigidity reached by the dominant UHECR sources lies well below the threshold for pion production on CMB photons (see, e.g., \citet{Allard2012}).

Soon after the Pierre Auger Observatory (hereafter Auger) first reported a gradual shift toward a heavier mass composition above the ankle, \citet{Decerprit2011} pointed out that a source population accelerating protons beyond $10^{20}$~eV, undergoing strong cosmological evolution, and contributing only subdominantly to the observed UHECR flux could nevertheless produce large cosmogenic neutrino fluxes. As an example, they considered the population of powerful FRII radio galaxies \citep{Fanaroff1974}, adopting the cosmological evolution proposed by \citet{Wall2005}. Assuming a proton maximum energy above $10^{20}$~eV, they showed that a 10\% contribution from such sources around $10^{19}$~eV would produce as many cosmogenic neutrinos as a pure-proton population following the star formation rate (SFR) evolution. Although these two scenarios are equivalent in terms of cosmogenic neutrino production, the subdominant-contribution scenario was not significantly challenged by the composition measurements reported by Auger. This opened the possibility that a source population making only a modest contribution to the present-day UHECR flux could nevertheless dominate the production of cosmogenic neutrinos.

The existence of such a subdominant population of UHECR accelerators reaching much higher maximum rigidities than the dominant sources appears quite natural. The maximum rigidity achievable at a given source is expected to scale with its magnetic luminosity, $L_{\rm B}$, which is itself generally related to the bolometric luminosity through simple equipartition or minimum-energy arguments. For instance, the estimate derived by \citet{Achterberg2002},
\[
E_{\rm max}\sim 2.5\times10^{20}Z\beta_{\rm s}\Gamma_{\rm s}
(L_{\rm B}/10^{46}\rm erg\,s^{-1})^{1/2},
\]
where $Z$ is the cosmic-ray charge, and $\beta_{\rm s}$ and $\Gamma_{\rm s}$ are the shock velocity and Lorentz factor, respectively (see also \citet{Levinson2006} for an alternative discussion), shows that accelerating protons beyond $10^{20}$~eV requires extremely luminous sources. Such objects are expected to be very rare in the nearby Universe (and within the Greisen--Zatsepin--Kuzmin horizon). FRII radio galaxies are among the most plausible candidates \citep{Rachen1993}, thanks to their large hot spots, where particle acceleration is known to occur in magnetic fields of order 100~$\mu$G. Their contribution to the diffuse UHECR flux at Earth is nevertheless expected to remain subdominant with respect to the much more numerous population of less luminous sources reaching maximum rigidities of only a few $10^{18}$~V, as required to reproduce the Auger spectrum and composition measurements.

More recently, the idea of a subdominant proton component was revisited by \citet{Globus2017}. They showed that such a population, undergoing strong cosmological evolution as discussed above, could be added to the dominant UHECR component (see Sect.~\ref{Sect:phot} for more details) without violating the \textit{Fermi}-LAT constraints on the diffuse extragalactic $\gamma$-ray background \citep{Ackermann2015, Ackermann2016}. As in the earlier work of \citet{Decerprit2011}, FRII radio galaxies were adopted as an illustrative example, contributing 5\% of the UHECR flux at $10^{19}$~eV.

The resulting cosmogenic neutrino flux above $10^{17}$~eV is much larger than that produced by the dominant UHECR population, owing to the combination of the strong cosmological evolution of FRII galaxies and their high maximum proton energy. Remarkably, the predicted flux closely matches that inferred from the KM3NeT observation of the ultra-high-energy neutrino event KM3-230213A \citep{KM3NeTobs2025, KM3NeTCorrected2025}, once the Auger \citep{Aab2019} and IceCube \citep{Aartsen2018} exposures in the same energy range are taken into account (see the discussion and Fig.~8 of \citet{Globus2025}).

It is of course extremely difficult to draw any firm conclusion from a single UHE neutrino event, given the large statistical uncertainty and the variety of astrophysical mechanisms that could potentially account for it. Nevertheless, the agreement between the flux predicted for the subdominant UHECR component associated with the strong cosmological evolution of bright FRII galaxies and that inferred from the KM3-230213A event is intriguing. 

The aim of the present work is to investigate whether FRII radio galaxies provide a physically motivated realization of this class of scenarios and to assess its compatibility with current multimessenger observational constraints. 
Unlike previous studies, we do not prescribe the contribution of FRII radio galaxies to the observed UHECR flux. Instead, the UHECR emissivity is derived self-consistently from the observed FRII luminosity function, combined with empirical radio--kinetic power relations and a physically motivated cosmic-ray efficiency. The resulting FRII contribution is therefore a prediction of the model rather than an imposed normalization.

We then investigate whether the resulting cosmogenic neutrino flux can still account for the KM3-230213A event and discuss the prospects for detecting both the diffuse flux and individual point sources with current and future neutrino observatories.

In Sect.~2, we describe our model for the luminosity-dependent density evolution of FRII galaxies and specify our assumptions regarding their UHECR output. We then confront this astrophysical scenario with the main multimessenger observables. Section~3 first examines its diffuse signatures, namely the UHECR, cosmogenic neutrino, and diffuse $\gamma$-ray fluxes, before discussing the expected signatures of individual FRII galaxies in both UHECRs and neutrinos. Finally, Sect.~4 summarizes the main conclusions of this exploratory study and discusses its implications for future observations.


\begin{figure*}
        \centering
        \includegraphics[width=8cm]{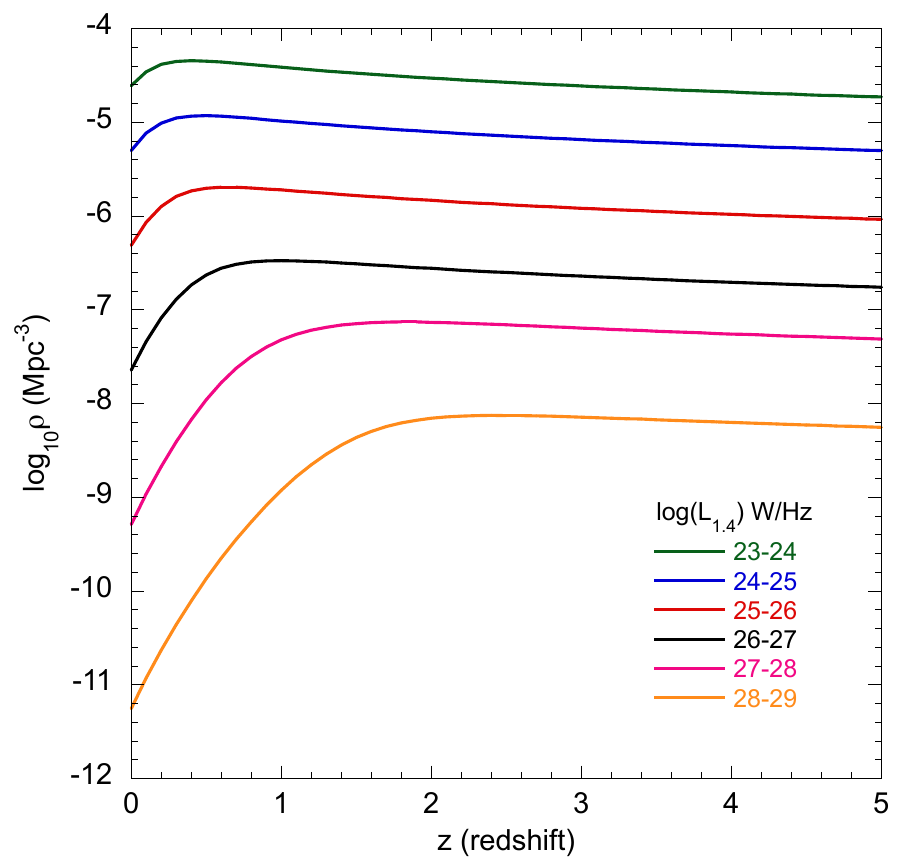}
        \includegraphics[width=8cm]
        {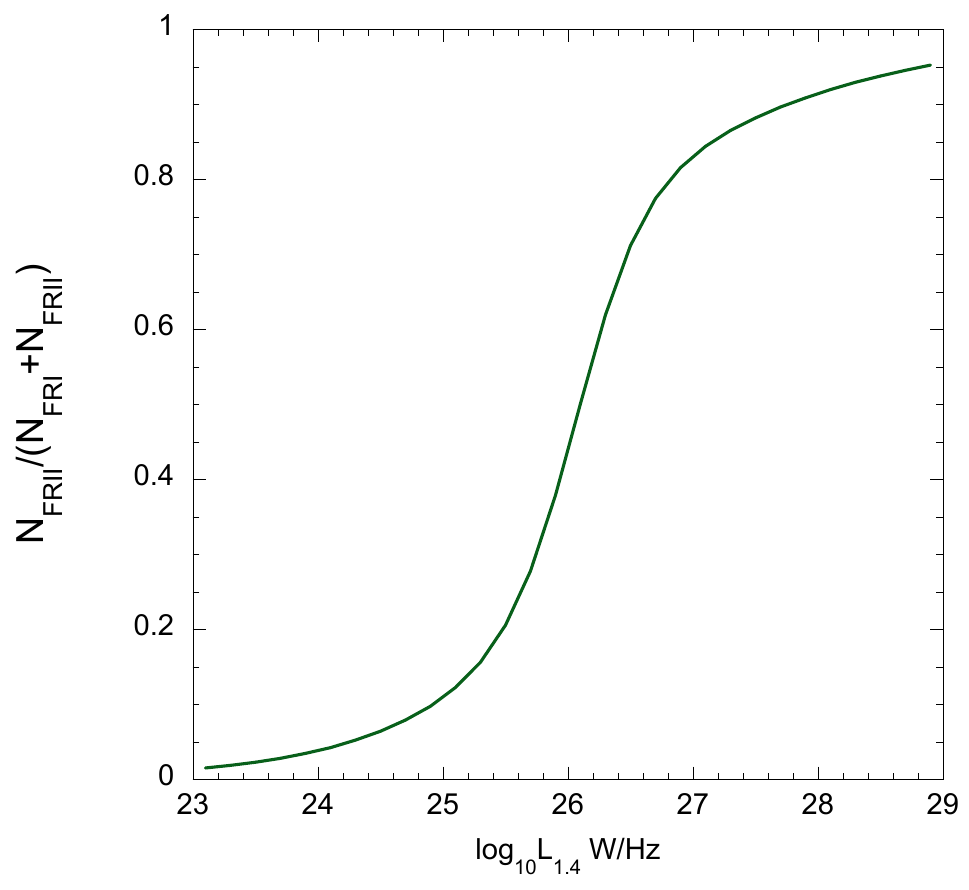}
        
        \caption{Left : Evolution of the comoving density of RGs as a function of the redshift for various $L_{1.4}$ luminosity bins as predicted by the LDDE of \citet{Slaus2024}. Right : Luminosity evolution of the fraction of FRII galaxies among all the RGs deduced from \cite{deJong2024}.}
        \label{Density}
\end{figure*}

\section{Model assumptions}
\label{ModelHypotheses}

\subsection{Luminosity-dependent density evolution of FRII galaxies}
\label{Sect:LDDE}

To describe the cosmological evolution of radio galaxies, we adopt the luminosity-dependent density evolution (LDDE) model proposed by \citet{Slaus2024}. Using a sample of more than 5400 radio galaxies (RGs) drawn from several radio surveys, the authors derived the luminosity function in the 1.4~GHz band and its cosmological evolution, testing several physically motivated functional forms. The best fit to the data was obtained with the following LDDE model:
\begin{equation}
\Phi(L,z)=\Phi_0(L),
\frac{(1+z_c)^{p_1}+(1+z_c)^{p_2}}
{\left(\frac{1+z_c}{1+z}\right)^{p_1}
+\left(\frac{1+z_c}{1+z}\right)^{p_2}},
\end{equation}
where the local luminosity function $\Phi_0(L)$ is given by
\begin{equation}
\Phi_0(L)=
\Phi^\star
\left(\frac{L}{L^\star}\right)^{1-\alpha}
\exp\left[
-\frac{1}{2\sigma^2}
\left(
\log\left(1+\frac{L}{L^\star}\right)
\right)^2
\right],
\end{equation}
and the characteristic redshift $z_c$ depends on luminosity according to
\begin{equation}
z_c=
\begin{cases}
z_c^\star, & L>L_a,\\[0.2cm]
z_c^\star \left(\frac{L}{L_a}\right)^a, & L\le L_a.
\end{cases}
\end{equation}

The best-fit values of the parameters $\Phi^\star$, $L^\star$, $\sigma$, $\alpha$, $z_c^\star$, $a$, $L_a$, $p_1$, and $p_2$ are reported in \citet{Slaus2024}. The left panel of Fig.~\ref{Density} shows the resulting redshift evolution of the comoving number density of RGs in several radio-luminosity bins. As also observed at other wavelengths, for instance in X-ray surveys (see, e.g., \citet{Hasinger2005}), the density evolution depends strongly on radio luminosity.

Once the LDDE model has been specified, we need an estimate of the fraction of FRII galaxies among RGs as a function of radio luminosity, since only FRII galaxies are considered throughout this work, except in Sect.~\ref{FRI}. Such an estimate was recently reported by \citet{deJong2024} (see also \citet{Gendre2013}). We use the local luminosity functions derived in that work for FRI and FRII galaxies\footnote{The summed local luminosity function of FRI and FRII galaxies estimated by \citet{deJong2024} is consistent with the local RG luminosity function derived by \citet{Slaus2024}.} to estimate the fraction of FRII galaxies as a function of 1.4~GHz radio luminosity.\footnote{Throughout this work, whenever necessary, radio luminosities are rescaled to the 1.4~GHz band using the estimated spectral index of each source, when available, or a value of 0.7 otherwise.} The resulting relation is shown in the right panel of Fig.~\ref{Density}. As is well known, FRII galaxies become increasingly dominant with increasing radio luminosity, with the transition between the FRI- and FRII-dominated regimes occurring around $L_{1.4}\simeq10^{26}\,\mathrm{W\,Hz^{-1}}$. Following \citet{deJong2024}, we assume that this fraction does not evolve with redshift.

Combining the FRII fraction with the LDDE model described above allows us to simulate different realizations of the FRII galaxy population as a function of luminosity and redshift. In the following subsection, we relate the 1.4~GHz radio luminosity, $L_{1.4}$, to the kinetic luminosity of the jet--lobe system, hereafter denoted by $L_{\rm kin}$, in order to estimate the potential UHECR output of FRII galaxies.

\subsection{Scaling relation between radio and kinetic luminosities in FRII galaxies}

For FRII radio galaxies, the relationship between radio luminosity ($L_{\rm radio}$) and jet kinetic luminosity ($L_{\rm kin}$) is commonly described by a power-law scaling relation of the form $L_{\rm kin}\propto L_{\rm radio}^{\beta}$, where $\beta$ typically lies in the range $0.8$--$0.9$. The widely used analytical model of \citet{Willott1999} predicts $\beta=6/7\simeq0.86$ based on minimum-energy arguments applied to radio lobes. Expressed in terms of the 1.4~GHz radio luminosity, the \citet{Willott1999} relation reads \citep{Smolcic2017}:
\begin{equation}
\log(L_{\rm kin})=0.86\log(L_{1.4})+21.08+1.5\log(f)
\label{Eq:Willott}
\end{equation}
where $L_{\rm kin}$ is expressed in erg\,s$^{-1}$, $L_{1.4}$ (the 1.4~GHz radio luminosity) in W\,Hz$^{-1}$, and $f$ is a parameter that accounts for several sources of uncertainty affecting the normalization of the relation, including the jet composition and possible departures from equipartition.

Many studies have attempted to estimate the kinetic luminosity of radio-galaxy jets. For RGs located in dense environments, this is often achieved from the properties of X-ray cavities (see, e.g., \citet{McNamara2007} for a review), or through dynamical modelling of the radio and/or X-ray spectral and morphological properties of the sources (see, e.g., \citet{Ito2008, Machalski2021}). For bright FRII galaxies, \citet{Daly2012} (but see also \citet{ODea2009}) derived a relation between $L_{\rm kin}$ and $L_{1.4}$ consistent with the prediction of \citet{Willott1999} for $f=4$, based on observations of 31 FRII radio galaxies.

More recently, \citet{Machalski2021} compiled an atlas of 361 FRII galaxies and estimated several physical quantities, including $L_{\rm kin}$. Throughout the remainder of this work, we adopt the power-law fit to the $L_{\rm kin}$--$L_{1.4}$ relation derived from this atlas:
\begin{equation}
\log(L_{\rm kin}) = 21.22 + 0.895\,\log(L_{1.4})
\label{Eq:Lkin}
\end{equation}

\begin{figure}
        \centering
        \includegraphics[width=9cm]{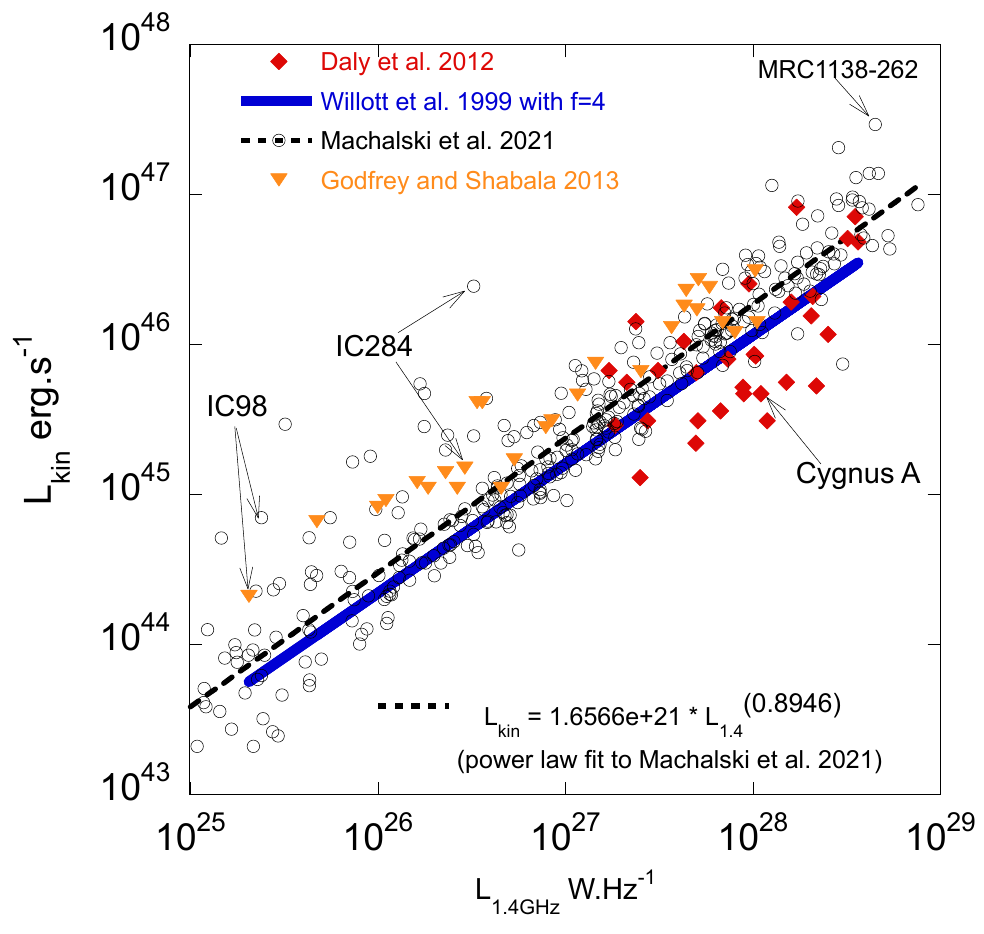}

        \caption{Relation between $L_{1.4}$ and $L_{\rm kin}$ using the data of \citet{Machalski2021, Daly2012, Godfrey2013}. The \citet{Willott1999} relation with $f=4$ is shown with a blue line, and the power law fit to \citet{Machalski2021} with a black dashed line. Some individual galaxies mentioned in the text (for which there might be several estimates) are shown with arrows.}
        \label{Lkin}
    \end{figure}

The slope and normalization of this relation are close to those predicted by the \citet{Willott1999} model with the value $f=4$ inferred by \citet{Daly2012}. This is illustrated in Fig.~\ref{Lkin}, where the data of \citet{Machalski2021}, \citet{Daly2012}, and \citet{Godfrey2013} are compared with the corresponding fitted relations. While the data of \citet{Machalski2021}, spanning more than five decades in $L_{1.4}$, clearly exhibit a power-law correlation between $L_{1.4}$ and $L_{\rm kin}$, the scatter around Eq.~\ref{Eq:Lkin} remains significant, amounting to approximately half a decade on either side of the best-fit relation.

Studies focusing mainly on lower-power FRI galaxies indicate that these objects are less radiatively efficient (see, e.g., \citet{Birzan2004, Birzan2008}). When interpreted within the \citet{Willott1999} framework, their estimated kinetic luminosities correspond to larger values of $f\simeq10$--15 (see the discussion in \citet{Smolcic2017, Ceraj2018} and references therein). This difference may partly reflect differences in jet composition between FRI and FRII galaxies (see below).

\subsection{Cosmic-ray output of FRII galaxies}
\label{UHECRoutput}

To complete our model, we now specify the three ingredients that determine the UHECR output of FRII galaxies: the cosmic-ray luminosity, the maximum proton energy, and the injected cosmic-ray spectrum.

\subsubsection{Cosmic-ray luminosity}

For the fractions of $L_{\rm kin}$ transferred to cosmic rays ($\epsilon_{\rm cr}$), electrons ($\epsilon_{\rm e}$), and magnetic fields ($\epsilon_{\rm B}$), we adopt the minimum-energy condition \citep{Burdidge1956, Miley1980}, which relates these three quantities through
$\epsilon_{\rm B}=\tfrac{3}{4}(1+k)\epsilon_{\rm e}$, with $k=\epsilon_{\rm cr}/\epsilon_{\rm e}$.

The total non-thermal energy is therefore shared between relativistic electrons, magnetic fields, and cosmic rays, $L_{\rm kin}=L_{\rm B}+L_{\rm e}+L_{\rm cr}$, which yields 
\begin{equation}
    \epsilon_{\rm cr} \equiv \frac{L_{\rm cr}}{L_{\rm kin}} = \frac{4}{7}\frac{k}{1+k}\,.
\end{equation} In the proton-dominated limit, $k\gg1$, this expression asymptotically approaches $ \epsilon_{\rm cr}\rightarrow\frac47$, so that $L_{\rm cr}=\epsilon_{\rm cr}L_{\rm kin}\simeq\frac47L_{\rm kin}$. 


The value of $k$, although highly uncertain, is expected to vary from one RG to another and is generally thought to be much larger in FRI galaxies, for which estimates of $k\gg100$ are often reported. Values inferred for FRII galaxies are generally lower, although still possibly well above unity (see, e.g., the discussions in \citet{Birzan2004, Birzan2008, Rafferty2006, Ito2008} and see e.g., \citet{Ineson2017} for a different conclusion). 
Consequently, our adopted value $\epsilon_{\rm cr}=4/7$ should be regarded as a fiducial optimistic normalization corresponding to the asymptotic proton-dominated limit, rather than as a direct prediction of equipartition theory. 
As noted above, all UHECR, neutrino and photon fluxes discussed below scale linearly with $\epsilon_{\rm cr}$.

\subsubsection{Maximum proton energy}
For the maximum proton energy reached in FRII acceleration regions, and in particular the FRII lobes considered here, we assume the standard scaling with the jet kinetic luminosity,
$E_{\rm max}\propto L_{\rm kin}^{1/2}$,
which follows from the Hillas confinement condition, $E_{\rm max}\propto BR$ ($R$ being the size of the acceleration site), together with the assumption that the magnetic luminosity is proportional to the jet kinetic power, $L_{\rm B}\propto B^2R^2c\propto L_{\rm kin}$.
We further define a reference kinetic luminosity,
$L_{\rm kin}^{\rm ref}$,
for which the maximum proton energy is
$E_{\rm max}^{\rm ref}$.
Throughout this work we adopt
$L_{\rm kin}^{\rm ref}=10^{47}\,\rm erg\,s^{-1}$,
so that

\begin{equation}
E_{\rm max}(L_{\rm kin})=
E_{\rm max}^{\rm ref}
\left(\frac{L_{\rm kin}}{10^{47}\rm erg\,s^{-1}}\right)^{1/2}.
\label{Eq:Emax}
\end{equation}

We consider three models, hereafter referred to as Model~1, Model~2, and Model~3, corresponding to $E_{\rm max}^{\rm ref}=3\,10^{19}$~eV,
$10^{20}$~eV, and $3\,10^{20}$~eV, respectively.

\subsubsection{Injected cosmic-ray spectrum}

We assume that particles are accelerated through non-relativistic diffusive shock acceleration in FRII lobes \citep{Blandford1978}. The injected cosmic-ray spectrum is therefore taken to follow a power law with an exponential cutoff above $E_{\rm max}$,

\[
\frac{dN}{dE}(E)\propto E^{-\beta}\exp\!\left(-\frac{E}{E_{\rm max}}\right).
\]

Throughout this work, we adopt the canonical value $\beta=2$.

Since our calculations only follow cosmic rays above $10^{17}$~eV, we introduce a correction factor, $\omega(E_{\rm max},\beta)$, to account for the fact that only a fraction of the total cosmic-ray luminosity is emitted above this energy:

\begin{equation}
\omega(E_{\rm max},\beta)\simeq
\frac{\int_{10^{17}}^{\infty}
E\,E^{-\beta}
\exp\left(-\frac{E}{E_{\rm max}}\right)\,dE}
{\int_{1\,\rm GeV}^{\infty}
E\,E^{-\beta}
\exp\left(-\frac{E}{E_{\rm max}}\right)\,dE}.
\label{Eq:omega_cr}
\end{equation}

For $E_{\rm max}=10^{18}$, $10^{19}$, and $10^{20}$~eV (assuming $\beta=2$), the corresponding values of $\omega(E_{\rm max},\beta)$ are approximately 0.09, 0.18, and 0.25, respectively.

Throughout this work, we assume a purely protonic UHECR composition. For the adopted spectral index $\beta=2$, however, the predicted cosmogenic neutrino flux depends much more strongly on the assumed maximum rigidity than on the exact mass composition.

\subsection{Realizations of the FRII distribution in the Universe}

For our simulations, we generated 1000 realizations of the distribution of active FRII galaxies in the Universe. In practice, we divide the Universe into comoving-distance bins (corresponding to redshift bins up to $z=6$) and the luminosity function into 35 logarithmically spaced luminosity bins ranging from $L_{1.4}=10^{23}\,\rm W\,Hz^{-1}$ to $L_{1.4}=10^{30}\,\rm W\,Hz^{-1}$ (with a width of 0.2 dex). The adopted LDDE model provides the expected number of FRII galaxies in each two-dimensional bin, from which the actual number of sources is randomly drawn for each realization.

The central value of each $L_{1.4}$ bin is converted into a kinetic luminosity using Eq.~\ref{Eq:Lkin}, and then into a maximum proton energy according to Eq.~\ref{Eq:Emax} for Models~1, 2, and 3.

Sources included in the \citet{Machalski2021} atlas, as well as a few well-known nearby FRII galaxies (including Cygnus~A), are not drawn randomly but are included in every realization. Whenever an estimate of $L_{\rm kin}$ is available in the literature, this value is adopted directly instead of being inferred from Eq.~\ref{Eq:Lkin}. For Cygnus~A, we adopt by default the value $L_{\rm kin}=4.7\times10^{45}\,\rm erg\,s^{-1}$ from \citet{Daly2012, Godfrey2013}, which is about four times lower than the value predicted by the scaling relation (see discussion below), although alternative estimates will also be considered. This corresponds to maximum proton energies ranging from $\sim6\times10^{18}$~eV (Model~1) to $\sim6\times10^{19}$~eV (Model~3).

To avoid unrealistically nearby simulated sources, we also use the recent FRII catalogue of \citet{Lao2024}, covering about 25\% of the sky and containing more than 45\,000 FRII galaxies, together with the radio-galaxy catalogues of \citet{vanVelzen2012} and \citet{Rachen2019}, both of which include nearby radio galaxies. For each luminosity bin, these catalogues are used to determine the minimum distance at which an FRII galaxy can be randomly drawn. This procedure reduces the realization-to-realization dispersion.

\subsection{Numerical tools}

For each distance and luminosity bin, the propagation of UHECR protons, the production of secondary neutrinos, photons and electrons, as well as the development of electromagnetic cascades, are computed using the numerical tools described by \citet{Decerprit2011}. Photomeson interactions are treated with the SOPHIA event generator \citep{Mucke2000}, while the extragalactic background light (EBL) and its cosmological evolution are modeled using the prescription of \citet{Gilmore2012} for the infrared, optical, and ultraviolet backgrounds.

The propagation calculations presented in this work do not include an extragalactic magnetic field (EGMF). Its possible impact, in particular through propagation delays affecting both UHECRs and the cosmogenic neutrinos associated with individual sources, is discussed in Appendix~\ref{Deflections}.

\section{Model predictions: constraints and implications}
\label{Results}

In this section, we examine the observational consequences of this FRII model and its compatibility with current multi-messenger constraints. We first examine its diffuse multi-messenger signatures (UHECRs, cosmogenic neutrinos, and diffuse $\gamma$ rays) before discussing the implications for individual sources.

\subsection{Diffuse fluxes}
\label{Sect:Diffuse}

All diffuse fluxes discussed below are averaged over 1000 realizations of the FRII distribution in the Universe.

\begin{figure*}[ht!]
        \centering
        
          \includegraphics[width=8cm]{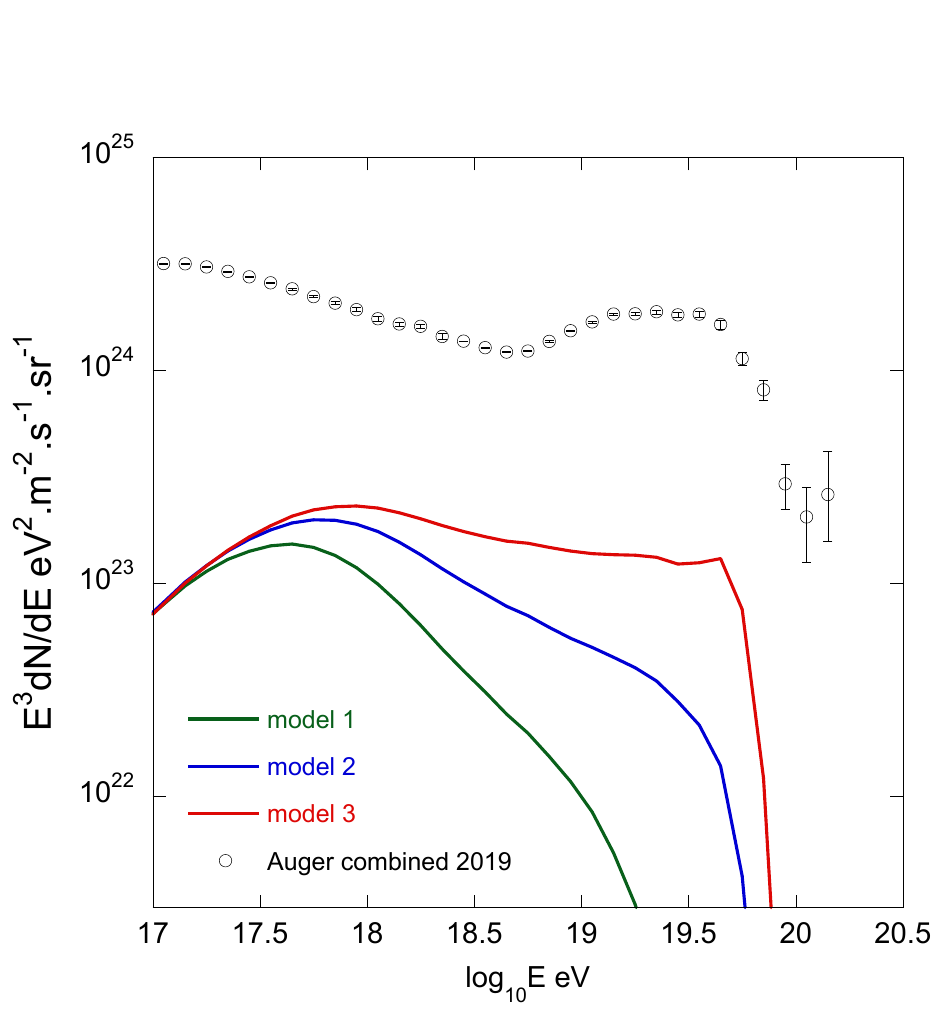}
        \includegraphics[width=8cm]
        {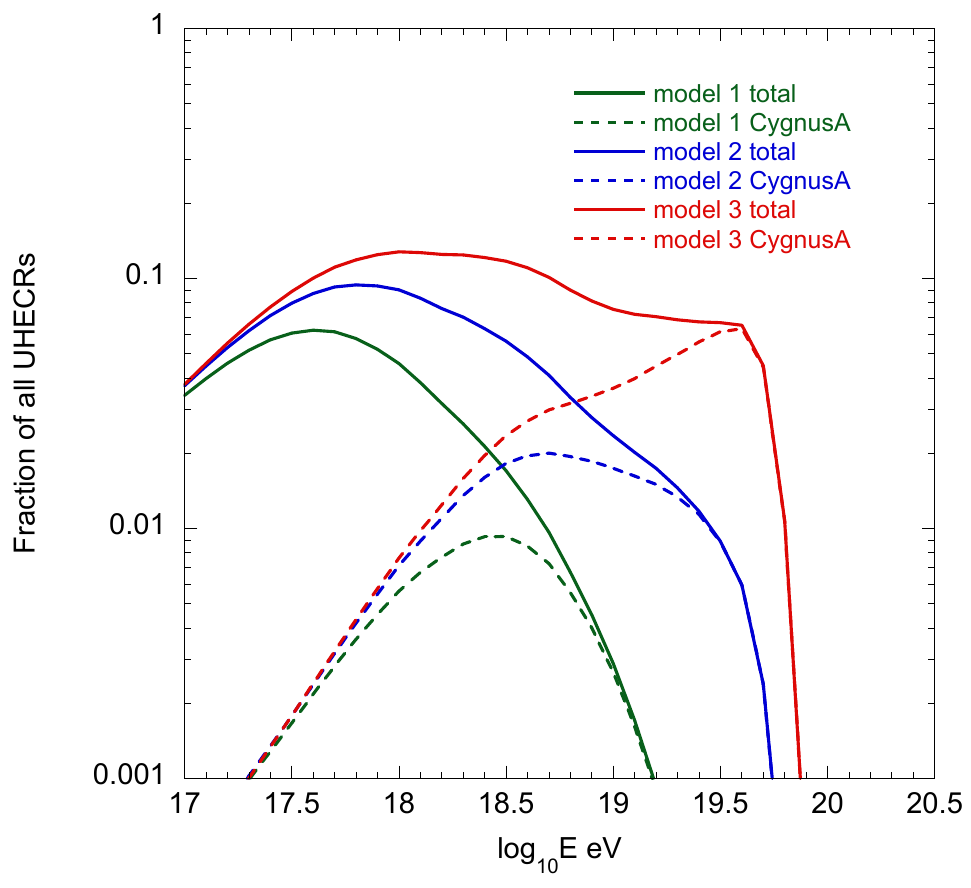}

        \caption{Left : UHECR diffuse spectra predicted for Models~1, 2, 3 compared with Auger data \citep{Verzi2019}. Right : Fraction of the total UHECR flux contributed by Models~1, 2 and 3. The combined contribution of all FRII galaxies is shown by solid lines, while that of Cygnus~A is shown by dashed lines.}
        \label{Neut_cr}
    \end{figure*}

\subsubsection{UHECRs}

The diffuse UHECR spectra, shown as $E^3dN/dE$, predicted for Models~1, 2, and 3 are compared with the Auger spectrum in the left panel of Fig.~\ref{Neut_cr}. The corresponding fractional contributions to the total UHECR flux are shown in the right panel. As expected, the contribution increases from Model~1 to Model~3 because of the increasing value of $E_{\rm max}^{\rm ref}$. However, regardless of the value adopted for $E_{\rm max}^{\rm ref}$, even if it exceeds that of Model~3, the contribution of FRII galaxies remains subdominant. In our calculations, it never exceeds approximately 12\%, reached by Model~3 around $10^{18}$~eV.

As noted above, these fluxes are computed assuming a negligible EGMF. Even if the EGMF in the local Universe were of order a few nG, strong magnetic-horizon effects capable of significantly modifying the diffuse UHECR spectrum around the ankle appear unlikely. This conclusion nevertheless deserves some caution. Although FRII jet--lobe systems are long-lived (typically a few $10^6$ to a few $10^7$~yr), they are transient rather than permanent sources and are also rare in the local Universe. If the EGMF were strong enough to prevent low-energy UHECRs from currently active nearby FRIIs from reaching Earth, the FRII contribution to the observed UHECR flux would not necessarily trace the currently active FRII population. However, the missing contribution from currently active FRIIs could in principle be compensated by delayed UHECRs emitted by nearby FRII galaxies that have since become inactive, or during previous activity cycles of the same galaxies. Although a quantitative assessment would require dedicated simulations, this suggests that magnetic-horizon effects may not strongly distort the diffuse FRII contribution to the UHECR spectrum. At energies above $10^{19}$~eV, such effects are expected to be even smaller, as discussed in Appendix~\ref{Deflections} in connection with the potential contribution of Cygnus~A to UHECR anisotropies.

Above $10^{19}$~eV, Model~3 contributes approximately 7\% of the total UHECR flux, consistent with the latest constraints on the composition from the Auger data \citep{Tkachenko2025}. For all three models, the relative contribution of Cygnus~A increases markedly toward the highest energies, as shown in the right panel of Fig.~\ref{Neut_cr}. In Model~3, Cygnus~A accounts for approximately 3\% of the total UHECR flux above $10^{19}$~eV and approximately 6\% above $3\times10^{19}$~eV. Consequently, the assumed values of $L_{\rm kin}$ and $E_{\rm max}$ for Cygnus~A have a strong impact on both the predicted FRII contribution to the UHECR spectrum and the expected anisotropy at the highest energies. We return to this issue in Sect.~\ref{UHECRpoint}, where we explore several estimates of $L_{\rm kin}$ from the literature together with the possible impact of extragalactic and Galactic magnetic fields.

\subsubsection{Cosmogenic neutrinos}

\begin{figure*}[ht!]
        \centering
         
       \includegraphics[width=8cm]{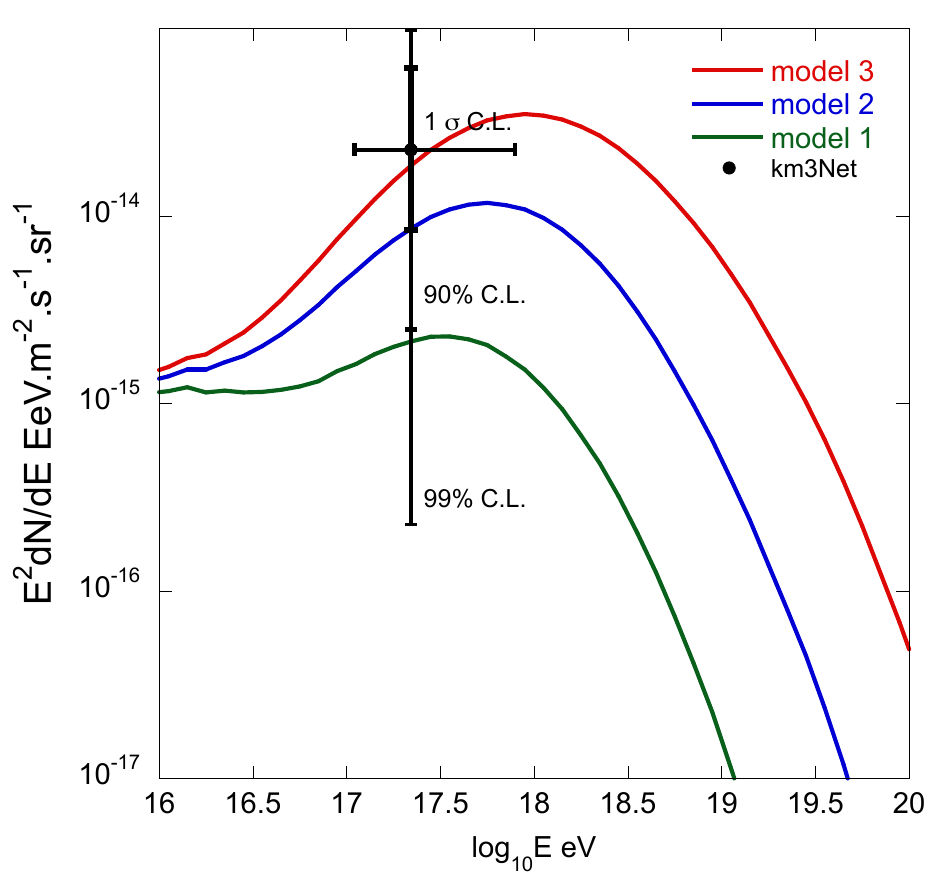}
        \includegraphics[width=8cm]
        {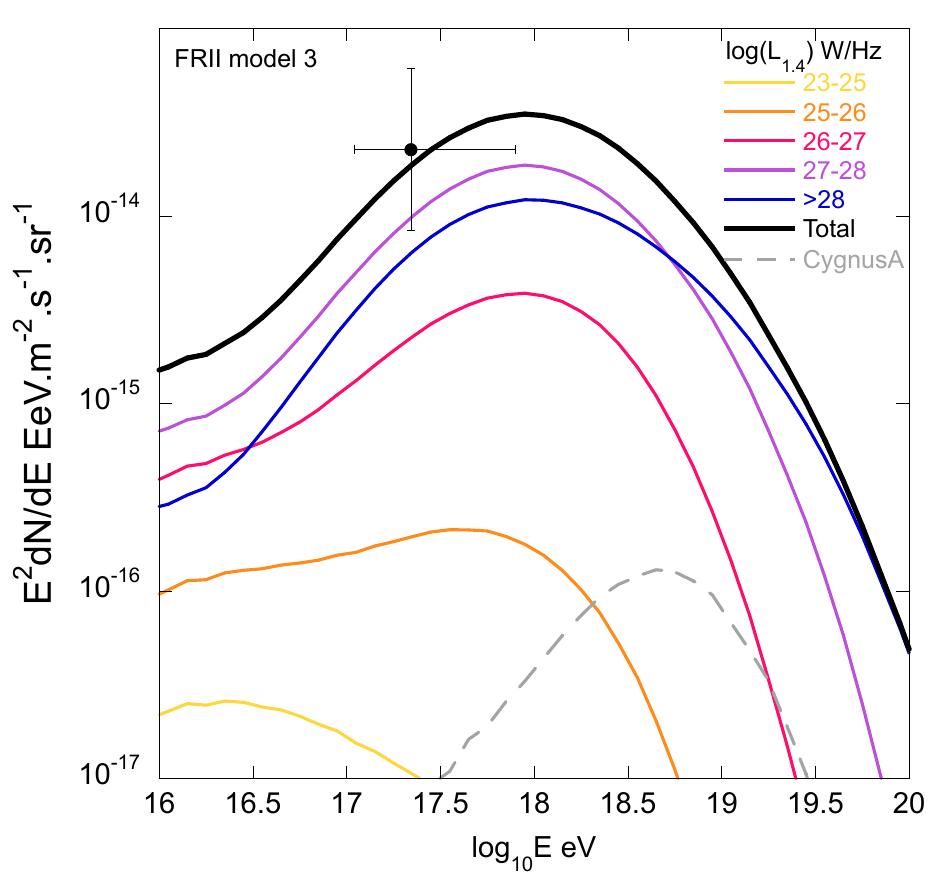}
        \includegraphics[width=8cm]
        {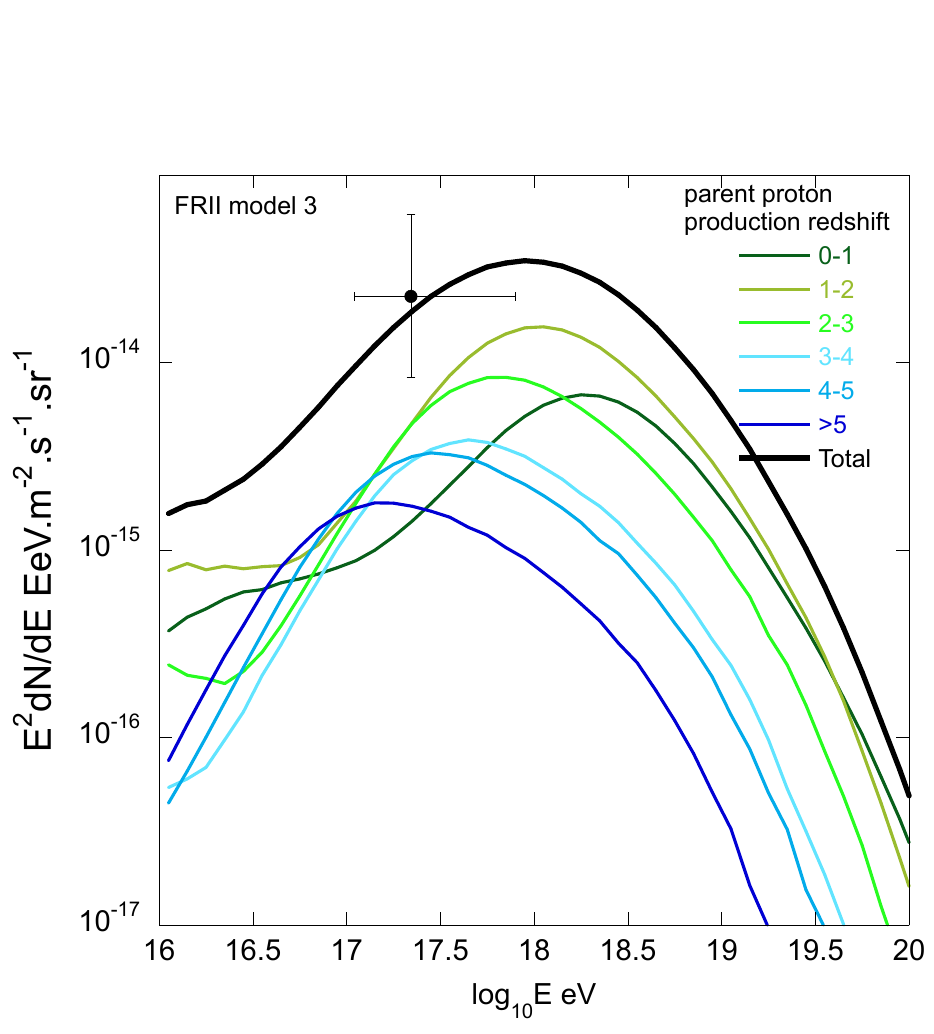}
        \includegraphics[width=8cm]{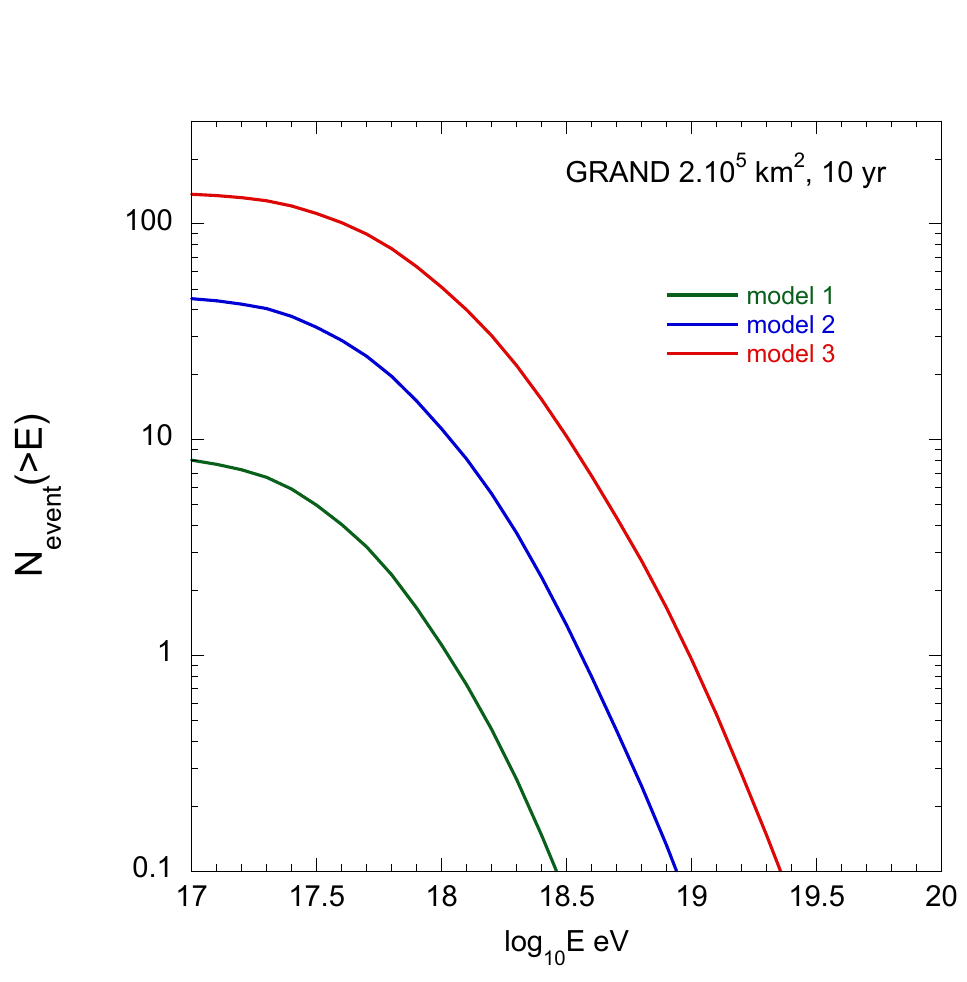}
        \caption{Diffuse neutrino flux predictions for our models. Top left: diffuse neutrino fluxes predicted for models 1, 2 and 3 and compared to the flux deduced from the KM3-230213A event \citep{KM3NeTCorrected2025}. Top right: Contribution of various $L_{1.4}$ bins to the diffuse neutrino flux predicted for model 3, the contribution of Cygnus A is also shown. Bottom left: Contribution of various redshift bins to the diffuse neutrino flux predicted for model 3. Bottom right: Number of events expected after 10 years of observations by GRAND assuming a total array area of $2\,10^5\rm\,km^2$ as implied by the diffuse fluxes predicted for models 1, 2 and 3.}
        \label{Neut_contrib}
    \end{figure*}

The diffuse cosmogenic neutrino fluxes ($E^2dN/dE$) predicted for Models~1, 2, and 3 are shown in the top-left panel of Fig.~\ref{Neut_contrib} and compared with the flux inferred from the KM3-230213A event, as estimated by \citet{KM3NeTCorrected2025}. The corresponding $1\sigma$, 90\%, and 99\% confidence intervals, derived using the prescription of \citet{Feldman1998}, are also shown. Models~2 and 3 lie within the $1\sigma$ confidence interval, whereas Model~1 falls slightly below the lower boundary of the 90\% confidence interval. For comparison, a model with $E_{\rm max}^{\rm ref}=10^{21}$~eV would predict approximately 50\% more neutrinos at $2\times10^{17}$~eV and roughly a factor of two more at the peak than Model~3. Models with $E_{\rm max}^{\rm ref}\ge10^{20}$~eV therefore appear to provide cosmogenic neutrino fluxes compatible with the current observational picture, namely the flux inferred from KM3-230213A together with the non-detection of UHE neutrinos by Auger and IceCube.

The top-right panel of Fig.~\ref{Neut_contrib} shows, for Model~3, the contributions of different FRII radio-luminosity intervals to the diffuse cosmogenic neutrino flux. 
The flux is largely dominated by FRIIs with $L_{1.4}\geq10^{27}\,\rm W\,Hz^{-1}$, which both accelerate UHECRs to higher maximum energies and exhibit a much stronger cosmological evolution than lower-luminosity FRIIs.

The contribution of Cygnus~A, one of the brightest individual sources in our model, is also shown. As discussed in more detail in Sect.~\ref{Neutpoint}, its contribution to the diffuse neutrino flux remains much smaller than its contribution to the UHECR flux, never exceeding approximately 1\% of the total neutrino flux for the adopted value of $L_{\rm kin}$.

The bottom-left panel of Fig.~\ref{Neut_contrib} shows the contributions from different redshift intervals. Near the energy of the KM3-230213A event, the dominant contribution arises from sources between $z=1$ and $z=3$, which account for approximately 50\% of the diffuse flux at $2\times10^{17}$~eV. Sources beyond $z=4$, where the LDDE is less well constrained (see the discussion in \citet{Slaus2024}), still contribute about 25\% of the flux at the same energy, although their relative contribution decreases toward higher neutrino energies.

The predicted diffuse fluxes can be converted into expected numbers of detected events using the effective areas of current and planned neutrino observatories. We performed these calculations for GRAND \citep{GRAND2020} and KM3NeT \citep{Adrian2016}. For the completed KM3NeT detector, we estimated the effective area by scaling that of its current partial configuration corresponding to data selection cuts for highest energy events, as reported by \citet{KM3NeTCorrected2025}, to the final instrumented volume. For Model~3, we predict about 2.5 events above $10^{17}$~eV (0.6 above $10^{18}$~eV) over 10 years of observations with the completed KM3NeT detector. Such low event rates would imply that the detection of KM3-230213A at the current stage of the detector corresponds to a positive statistical fluctuation. This interpretation is, however, already suggested by the combination of the neutrino flux inferred from KM3-230213A with the non-detection of UHE neutrinos by Auger and IceCube.

The corresponding expectations for GRAND are shown in the bottom-right panel of Fig.~\ref{Neut_contrib}, assuming 10 years of observations and a total instrumented area of $200,000\,\rm km^2$. The planned GRAND200k observatory is expected to consist of twenty $10^4\,\rm km^2$ sub-arrays distributed worldwide, whose effective areas were estimated in \citet{GRAND2020}. Throughout this paper, we adopt the most optimistic trigger strategy considered in that work. For Models~2 and~3, the expected number of detected neutrinos above $10^{17}$~eV ranges from about 50 to 135 over 10 years. Above $10^{18}$~eV, the corresponding range is about 12--50 events, decreasing to roughly 1.5--10 events above $10^{18.5}$~eV. Such statistics should be sufficient to constrain the overall shape of the neutrino spectrum. Model~1 predicts substantially fewer events and is, among the three models considered, the least compatible with the constraints derived from the combined KM3NeT, Auger and IceCube data. Although measuring the diffuse spectrum would already provide valuable constraints on the origin of UHE neutrinos, identifying the underlying astrophysical sources will ultimately require the detection of anisotropies and, ideally, individual point sources.


\begin{figure*}
        \centering
        \includegraphics[width=8cm]{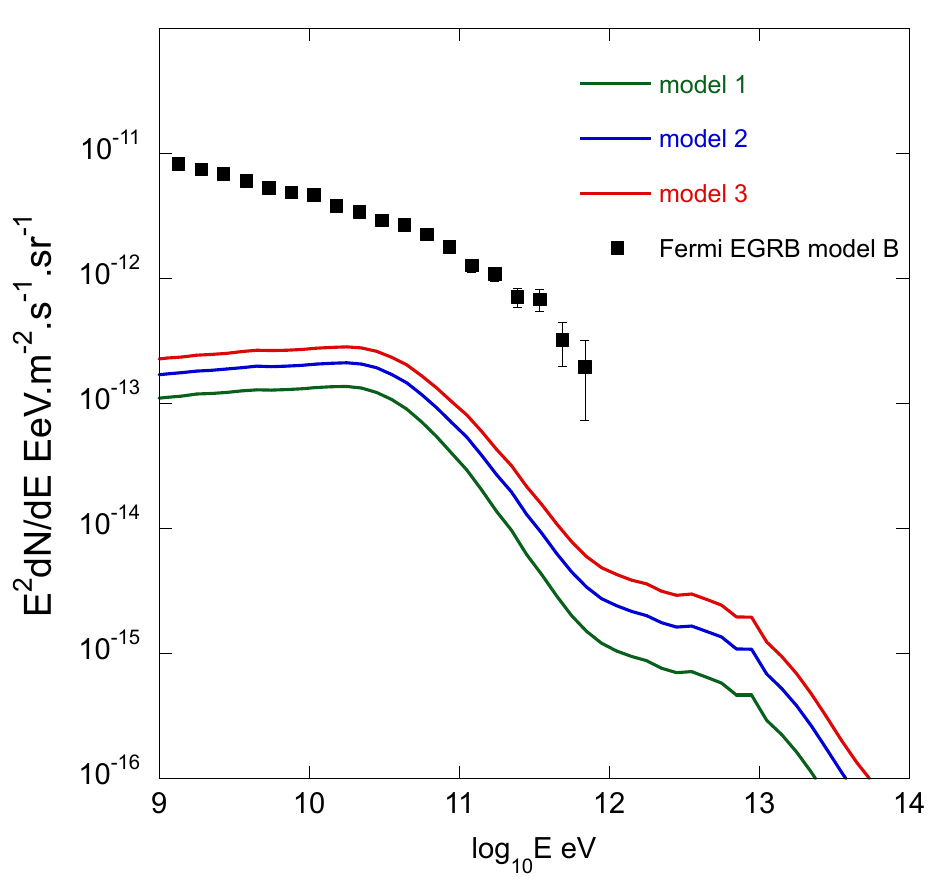}
        \includegraphics[width=8cm]
        {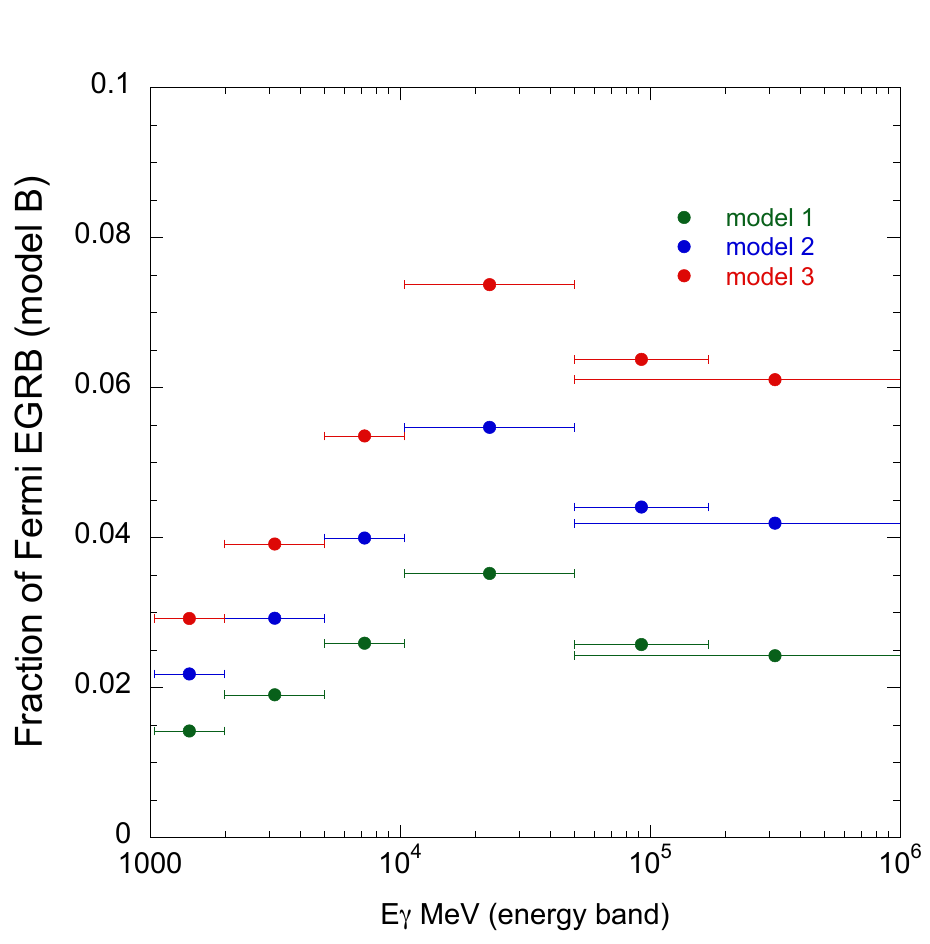}

        \caption{Left : diffuse photon fluxes predicted for models 1, 2 and 3 compared to EGRB (model B) estimated from Fermi-LAT data \citep{Ackermann2015}. Right : fractional contribution to the EGRB (model B) of the photon fluxes predicted for models 1, 2 and 3.}
        \label{Phot}
    \end{figure*}

    \begingroup

\setlength{\tabcolsep}{10pt} 
\renewcommand{\arraystretch}{1.7} 
\begin{table*}
\caption{Contributions to the EGRB: $F_{\rm PS}$ refers to the flux from resolved and unresolved point sources as estimated by \citet{Ackermann2016, Zechlin2016}, $F_{\rm SFG+misAGN}$ corresponds to the summed contribution of SFG and misAGN for which we consider the models of \citet{Ackermann2012, Inoue2011}, $F_{\rm UHECR}^{\rm main\,GRB}$ corresponds to the main UHECR component considered in \citet{Globus2017} assuming a GRB-like source evolution and $F_{\rm UHECR}^{\rm sub\,FRII}$ refers to the cosmogenic photon flux calculated for model 3.}                 
\label{table:egrb}    
\centering                        
\begin{tabular}{c c c c c c c}      
\hline\hline  
\textbf{Energy bands} (in GeV) & 1.04--1.99 & 1.99--5.0 & 5.0--10.4 & 10.4--50 & 50--171 & 50--2000 \\        

\hline\hline
  $F_{\rm PS}/F_{\rm EGRB}$ (\% Model B) & $68^{+5}_{-10}$ & $63^{+4}_{-13}$ & $52^{+15}_{-6}$ & $51^{+22}_{-4}$ & $65^{+41}_{-15}$ & $71^{+13}_{-12}$ \\    
  \hline
  $F_{\rm SFG+misAGN}/F_{\rm EGRB}$ (\% Model B) & $25^{+27}_{-10}$ & $23^{+25}_{-9}$ & $20^{+23}_{-8}$ & $16^{+20}_{-7}$ & $6^{+7}_{-3}$ & $6^{+6}_{-2}$ \\
  \hline
  $F_{\rm UHECR}^{\rm main\,GRB}/F_{\rm EGRB}$ (\%  Model B) & 7.0 & 9.5 & 13 & 17 & 13 & 13 \\
\hline  
 $F_{\rm UHECR}^{\rm sub\,FRII}/F_{\rm EGRB}$ (\% \rm Model B) & 2.9 & 3.9 & 5.3 & 7.3 & 6.4 & 6.1 \\
 \hline  
 $F_{\rm Total}/F_{\rm EGRB}$ (\% \rm Model B) & \textbf{103} & 99 & 90 & 91 & 90 & 96 \\
\end{tabular}
\end{table*}
\endgroup

\subsubsection{Cosmogenic photons}
\label{Sect:phot}

Like neutrinos, cosmogenic photons \citep{Strong1973, Strong1974, Berezinsky1975} are produced during UHECR propagation. Interactions of UHECR protons and nuclei with background photons generate high-energy photons as well as electrons and positrons, which subsequently initiate electromagnetic cascades. The diffuse cosmogenic photon fluxes predicted for Models~1, 2 and~3 are shown in the left panel of Fig.~\ref{Phot} and compared with the extragalactic gamma-ray background (EGRB) measured by Fermi-LAT \citep{Ackermann2015}, adopting Galactic foreground model~B\footnote{Estimating the EGRB requires subtracting the Galactic foreground emission, which must itself be modeled. The Fermi-LAT collaboration proposed three foreground models (A, B and C) based on different physical assumptions. Following \citet{Globus2017}, we adopt model~B, which yields the lowest Galactic foreground and therefore the highest EGRB estimate, providing the most conservative benchmark for gamma-ray constraints on UHECR models.}.

Unlike the cosmogenic neutrino fluxes, the predicted photon fluxes differ by less than a factor of two between the three models. This weaker dependence on $E_{\rm max}$ arises because cosmogenic photon production is dominated by Bethe--Heitler pair production on the CMB, which becomes important for protons around $10^{18}$~eV at $z=0$, and therefore does not require the highest source energies. This dependence on $E_{\rm max}$ is even weaker for soft source spectra, for which the photon production is more strongly dominated by particles in the energy range where pair production becomes important, a range populated in all three models. The corresponding fractional contributions of the predicted cosmogenic photon fluxes to the Fermi-LAT EGRB in the six energy bands are shown in the right panel of Fig.~\ref{Phot}.

The compatibility of these predictions with gamma-ray constraints cannot be assessed by simply comparing the predicted cosmogenic photon fluxes with the measured EGRB  since the EGRB itself contains several astrophysical components whose contributions must be taken into account.

First, the contribution of resolved and unresolved point sources, believed to be largely dominated by BL~Lac objects, has been estimated down to fluxes well below the Fermi-LAT detection threshold using the method proposed by \citet{Malyshev2011}. Estimates are available in six energy bands between 1~GeV and 2~TeV \citep{Ackermann2016, Zechlin2016, Lisanti2016}. In addition, as argued in \citet{Ackermann2016}, these estimates most likely do not include the contribution of faint gamma-ray emitters such as star-forming galaxies (SFGs) and misaligned AGNs (misAGNs), whose expected contributions should therefore also be included.

The dominant UHECR component must likewise be taken into account. Unlike the cosmogenic neutrino flux, its cosmogenic photon counterpart is not expected to be much smaller than that associated with the FRII population, owing to the weaker dependence of photon production on the maximum source energy. In fact, the reference model adopted by \citet{Globus2017}, inherited from the Galactic-to-extragalactic transition model of \citet{Globus2015}, was deliberately chosen to maximize the cosmogenic photon yield. It assumes both a soft proton spectrum at low energy, resulting from the source escape mechanism and allowing a good reproduction of the light-ankle feature reported by KASCADE-Grande \citep{Apel2013}, and a strong cosmological evolution of the source population following either the star-formation rate (SFR) or the gamma-ray burst (GRB) rate. Both ingredients increase the expected cosmogenic photon flux compared with harder source spectra or weaker source evolution.

Table~\ref{table:egrb} summarizes the various contributions to the EGRB considered by \citet{Globus2017}, assuming a GRB-like evolution for the dominant UHECR component, together with the additional contribution predicted for Model~3. The comparison is made in the same six energy bands in which the unresolved point-source contribution was estimated by \citet{Ackermann2016, Zechlin2016}.  The contribution of Model~3 is significantly smaller than that of the dominant UHECR component. Moreover, except for the lowest energy band, the sum of all the considered contributions remains below the measured EGRB. In the 1.04--1.99~GeV band, the combined flux exceeds the measured EGRB by only about 3\%, which is well within the uncertainties associated with the unresolved point-source contribution and with the poorly constrained contribution of faint gamma-ray populations (SFGs and misAGNs).

Overall, adding a subdominant FRII component to the UHECR population does not significantly modify the gamma-ray constraints on the origin of UHECRs or on the Galactic-to-extragalactic transition discussed by \citet{Globus2017}.

\subsection{Contribution and observability of individual sources}

\subsubsection{Cygnus A as a potential source of UHECR anisotropies}
\label{UHECRpoint}

Among individual FRII galaxies, Cygnus~A is expected to dominate the UHECR signal at the highest energies. We therefore focus primarily on this source when discussing the observability of individual FRII galaxies with UHECR observatories. Its exceptional role results from the combination of its very high radio luminosity, $L_{1.4}\simeq1.2\,10^{28}\,\rm W\,Hz^{-1}$, and its relatively small luminosity distance, $d_{\rm L}\simeq225$~Mpc.

Cygnus~A is located at the center of a rich galaxy cluster with an estimated mass of approximately $2.5\,10^{14}\,M_{\odot}$ \citep{Smith2002}. Such an environment is unusual for an FRII galaxy, since FRIIs are preferentially found in less dense environments than FRIs \citep[see, e.g.,][]{Prestage1988}. This peculiar environment may partly account for both its exceptional radio luminosity and its high inferred radiative efficiency, reflected in a comparatively low ratio of $L_{\rm kin}$ to $L_{1.4}$ (see Fig.~\ref{Lkin} and the discussions in \citet{Carilli1996, Godfrey2013, Birzan2004, Birzan2008}).

For completeness, we note that another nearby FRII galaxy could make a non-negligible contribution to the UHECR flux, namely IC98, although its contribution is expected to remain smaller than that of Cygnus~A. Located at a distance of approximately 125~Mpc, it is assigned a kinetic luminosity of $L_{\rm kin}\simeq7\,10^{44}\,\rm erg\,s^{-1}$ by \citet{Machalski2021}. However, \citet{Godfrey2013} derive a substantially lower estimate, $L_{\rm kin}\simeq2\,10^{44}\,\rm erg\,s^{-1}$, which nevertheless remains about three times larger than the value inferred from its radio luminosity using Eq.~\ref{Eq:Lkin}. Adopting this lower estimate reduces the expected contribution of IC98 to a level comparable to that of another nearby FRII galaxy, Pictor~A. In the following, we therefore focus on Cygnus~A, whose predicted contribution is substantially larger.

Although the overall contribution of FRII galaxies to the diffuse UHECR flux remains only a few percent, the situation is quite different when considering individual sources. In the explored scenario, the UHECR contribution from Cygnus~A consists predominantly of protons at the highest energies. Owing to their large magnetic rigidity, these particles are expected to undergo much smaller magnetic deflections than the nuclei dominating the bulk of the observed UHECR flux. Consequently, even a modest contribution to the total UHECR intensity could generate a detectable anisotropy with the currently available Auger statistics.

For our reference Model~3 and the baseline value $L_{\rm kin}=4.7\,10^{45}\,\rm erg\,s^{-1}$, Cygnus~A contributes approximately 6\% of the total UHECR flux above $3\times10^{19}$~eV. In the latest Auger dataset published above 32~EeV \citep{AugerAniso2022}, comprising about 2600 events, this would correspond to roughly one hundred events. Whether such a contribution should already have produced a detectable anisotropy is therefore an important question for the proposed scenario.

The answer depends primarily on two uncertainties. First, could extragalactic and Galactic magnetic fields sufficiently delay or deflect the UHECRs from Cygnus~A to suppress an observable anisotropy? Second, how do the uncertainties on the kinetic luminosity of Cygnus~A translate into the expected proton contribution and anisotropy signal?

The main conclusions of the analysis presented in Appendix~\ref{Deflections} can be summarized as follows. Although extragalactic magnetic fields may produce substantial time delays at low rigidities, they leave the predicted flux above a few $10^{19}$~eV almost unchanged even for field strengths as large as a few nG. Likewise, the expected angular deflections remain relatively modest. Including the Galactic magnetic field, most protons above 32~EeV are still expected to arrive within about $10^\circ$--$20^\circ$ of the source direction, depending on the magnetic-field model, while magnetic lensing may even enhance the observed flux in the direction of Cygnus~A. These results suggest that, for Model~3 and our baseline value of $L_{\rm kin}$, Cygnus~A should remain a plausible candidate for producing a detectable anisotropy on angular scales below about $15^\circ$. Such a signal is less expected in Model~2, for which the contribution of Cygnus~A above $3\times10^{19}$~eV remains below 1\%.

\begin{figure}[ht!]
        \centering
         
        \includegraphics[width=8cm]
        {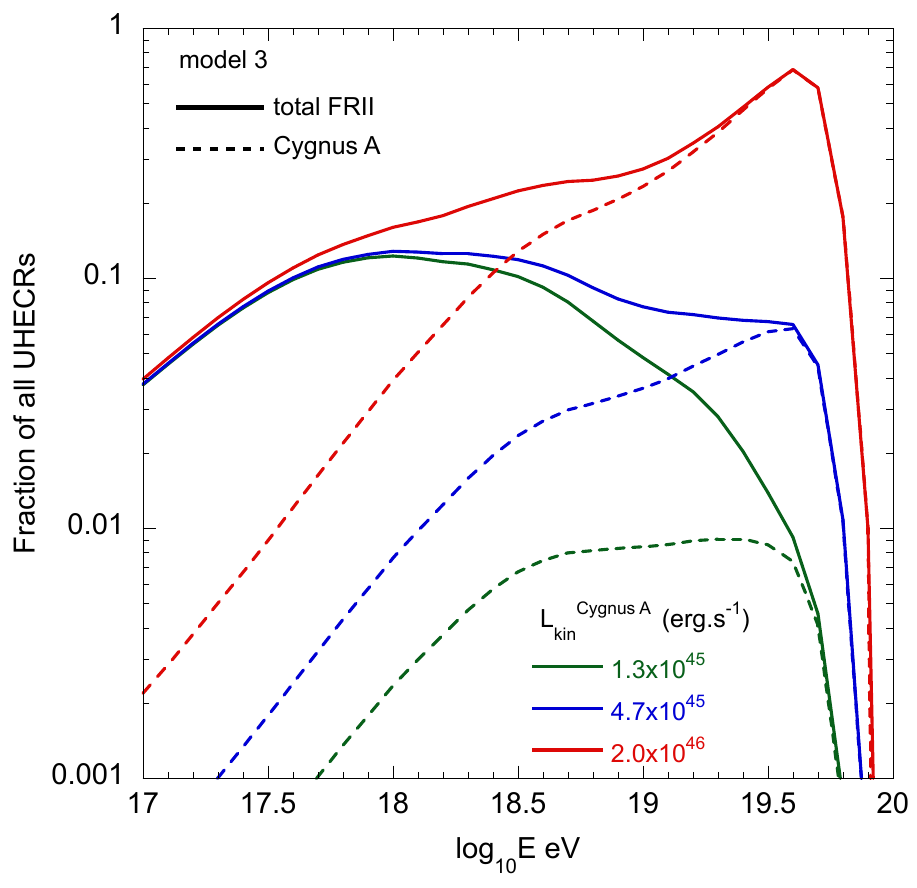}
        
    \caption{Fraction of the total UHECR flux predicted by Model~3 for three different values of $L_{\rm kin}$ for Cygnus~A (see legend and text). The combined contribution of all FRII galaxies is shown by solid lines, while the contribution of Cygnus~A is shown by dashed lines.}
        
        \label{CygA_crfrac}
    \end{figure}

\begin{figure*}[ht!]
        \centering
         
        \includegraphics[width=8cm]
        {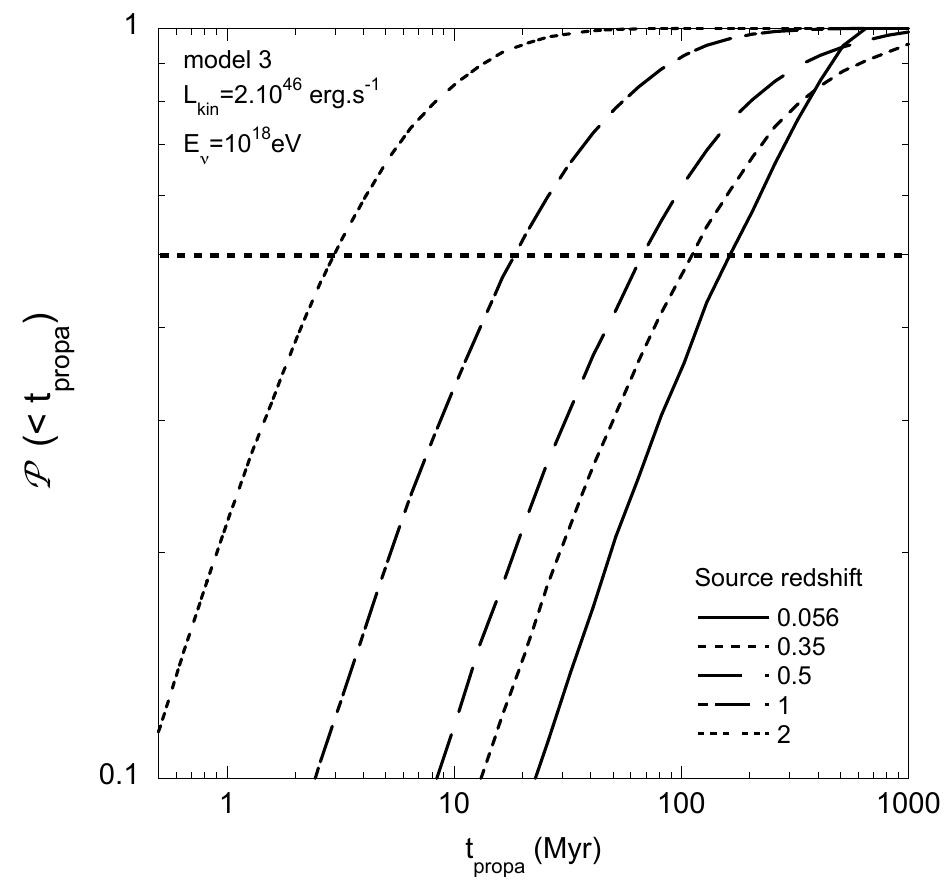}
         \includegraphics[width=8cm]
        {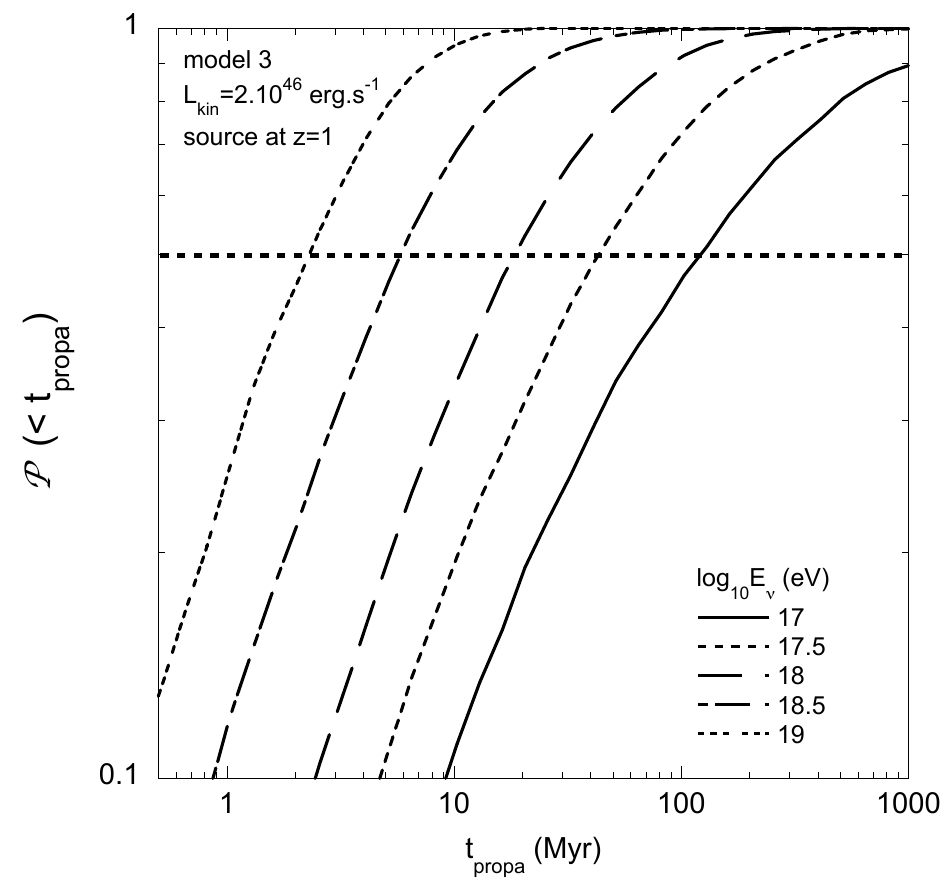}
        \caption{ Left : cumulative distribution of the elapsed time of UHECR proton propagation before producing a neutrino with energy $E_\nu=10^{18}$~eV for FRII galaxies (assuming $L_{\rm kin}=2\,10^{46}\rm\,erg\,s^{-1}$) located at various redshifs (see legend). Right :  cumulative distribution of the elapsed time of UHECR proton propagation before producing neutrinos of various energies (see legend) for a FRII galaxy (assuming $L_{\rm kin}=2\,10^{46}\rm\,erg\,s^{-1}$) located at $z=1$.}
        \label{time_prod}
    \end{figure*}

\begin{figure*}[ht!]
        \centering
         
        \includegraphics[width=8cm]
        {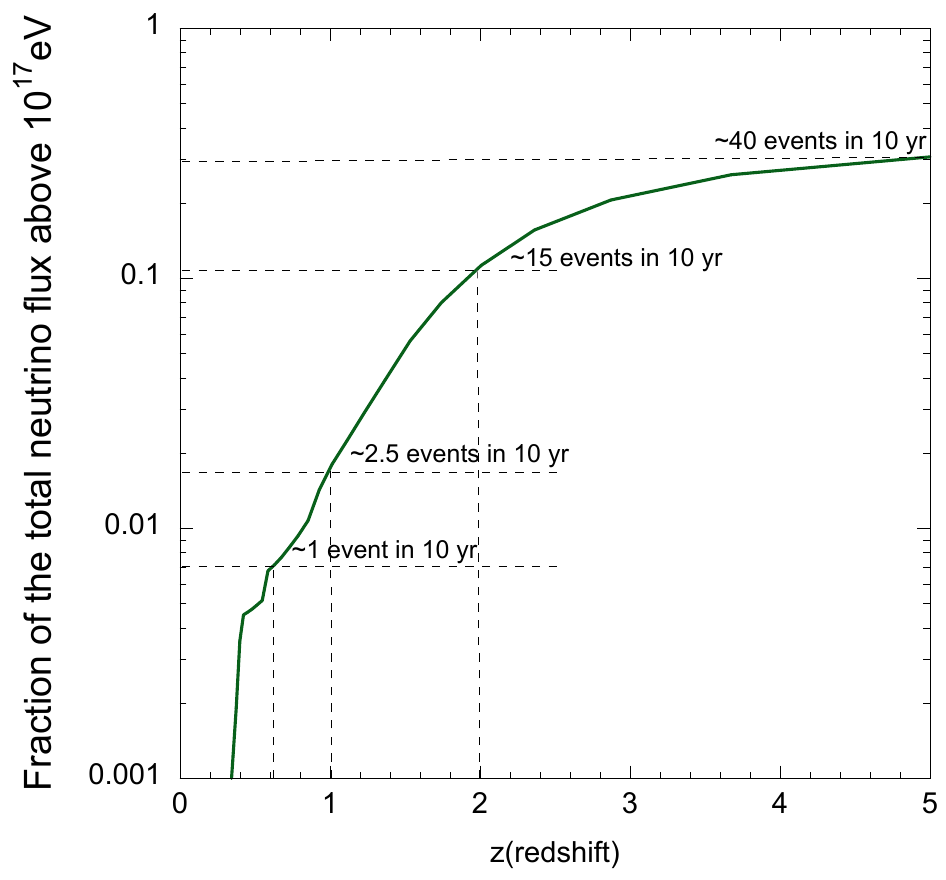}
        \includegraphics[width=8cm]
        {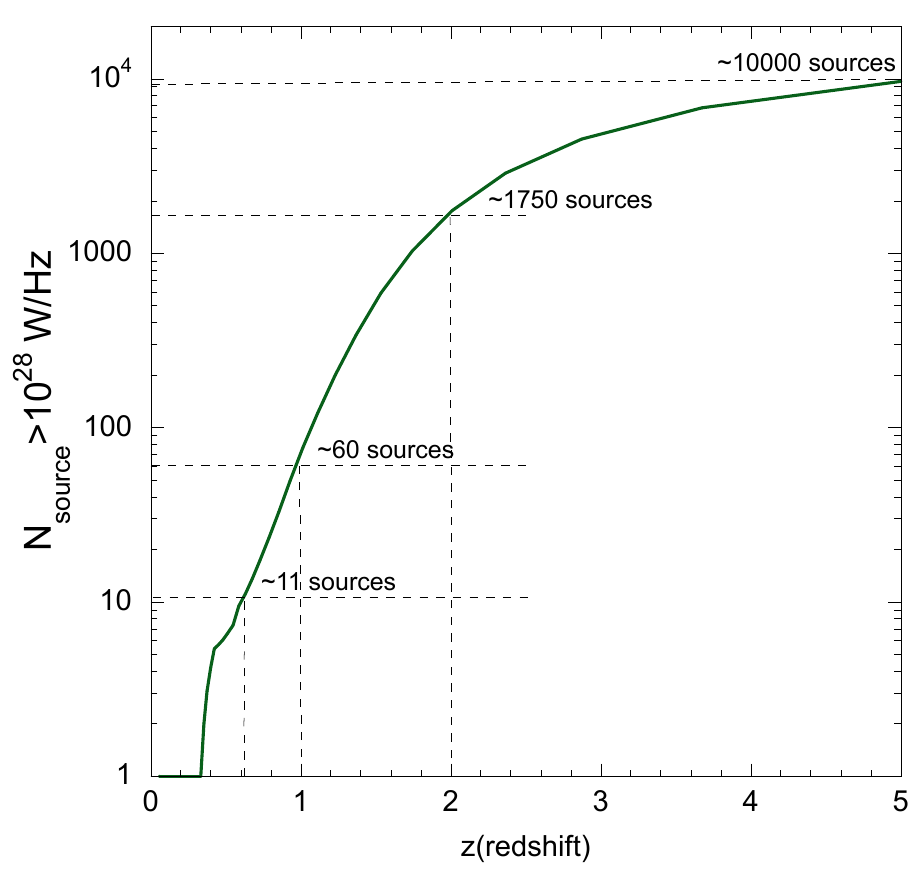}
        \caption{Left : relative contribution of FRII galaxies with $L_{1.4}\geq10^{28}\,\rm W\,Hz^{-1}$ to the diffuse neutrino flux of model 3 as a function of redshift. The corresponding number of events detected by GRAND in 10 yrs is indicated. Right : Number of FRII galaxies with $L_{1.4}\geq10^{28}\,\rm W\,Hz^{-1}$ enclosed as a function of redshift.}
        \label{fraction_28}
\end{figure*}

The second major uncertainty is the kinetic luminosity of Cygnus~A. It directly affects both the predicted anisotropy signal and the proton contribution at the highest energies, the latter being constrained by UHECR composition measurements \citep[see, e.g.,][]{Aab2014b}. Various estimates of $L_{\rm kin}$ have been proposed for Cygnus~A, based either on X-ray cavity measurements or on dynamical modelling of the radio lobes, spanning approximately one order of magnitude.

In addition to our baseline value $L_{\rm kin}=4.7\,10^{45}\,\rm erg\,s^{-1}$ \citep{Daly2012, Godfrey2013}, we therefore consider a lower estimate, $L_{\rm kin}=1.3\,10^{45}\,\rm erg\,s^{-1}$ \citep{Rafferty2006, Birzan2008}, and a higher value, $L_{\rm kin}=2\times10^{46}\,\rm erg\,s^{-1}$, close to the upper end of the interval estimated by \citet{Ito2008} and also close to the value inferred from Eq.~\ref{Eq:Lkin}. For these two values we set $E_{\rm max}$ according to the assumptions of Model~3 and recomputed the corresponding UHECR fluxes.

As shown in Fig.~\ref{CygA_crfrac}, the assumed value of $L_{\rm kin}^{\rm CygnusA}$ has a major impact on the expected contribution of Cygnus~A. For $L_{\rm kin}^{\rm CygnusA}=2\times10^{46}\,\rm erg\,s^{-1}$, the contribution reaches about 60\% around $10^{19.5}$~eV, implying a dominant proton component at these energies. Even if this component were completely isotropized by cosmic magnetic fields, such a large proton fraction above $10^{19}$~eV would be difficult to reconcile with Auger composition measurements \citep[see the latest analysis by][]{Tkachenko2025}. Conversely, for $L_{\rm kin}^{\rm CygnusA}=1.3\,10^{45}\,\rm erg\,s^{-1}$, the contribution of Cygnus~A decreases to about 0.8\%, while the total FRII contribution is about 1.4\%. At such a low level, the absence of a detectable anisotropy around Cygnus~A would be much easier to reconcile with the FRII scenario.



\subsubsection{Observability of cosmogenic neutrino point sources}
\label{Neutpoint}

Before discussing the detectability of cosmogenic neutrino point sources, one should first verify that cosmogenic neutrinos remain sufficiently well correlated with their parent FRII galaxies. This requires that the parent protons undergo photomeson interactions with the CMB photons (leading to the emission of secondary neutrinos) sufficiently rapidly after leaving the source, before magnetic deflections significantly alter their propagation direction. Since these parent protons mostly have energies above $10^{19}$~eV, magnetic deflections are themselves expected to remain modest (see the previous subsection). Both effects therefore contribute to preserving a close angular correlation between the neutrinos and their parent FRII galaxies.

Figure~\ref{time_prod} quantifies the propagation time of the parent protons before the emission of cosmogenic neutrinos. The left panel shows the cumulative production time of $10^{18}$~eV neutrinos for FRII galaxies with $L_{\rm kin}=2\times10^{46}\,\rm erg\,s^{-1}$ at various redshifts: $z=0.056$ (Cygnus~A), $0.35$ (the redshift of the closest FRII galaxy in the \citet{Lao2024} catalog expected to have this value of $L_{\rm kin}$ according to Eq.~\ref{Eq:Lkin}), $0.5$, $1$, and $2$. The right panel shows the cumulative production time for neutrinos of various energies from an FRII galaxy with the same luminosity at $z=1$.

Neutrino production becomes more efficient at higher redshift owing to the increasing CMB energy density.
Above $z\sim1$, where most cosmogenic neutrinos are produced (see Fig.~\ref{Neut_contrib}), essentially all $10^{18}$~eV neutrinos originate within $\sim30$~Mpc of the proton source. These calculations therefore support the assumption adopted throughout the remainder of this section that cosmogenic neutrinos remain tightly correlated with their parent FRII galaxies.

Under the above assumption, we can now investigate the detectability of individual cosmogenic neutrino point sources. We calculated the number of neutrinos that GRAND ($2\,10^{5}\,\rm km^2$ and 10 years of observation time) would detect from the FRII galaxies of the \citet{Machalski2021} atlas and those compiled by \citet{Daly2012}, assuming models~2 and~3. For these calculations, we assumed that the full 200k GRAND observatory consists of 20 sites whose locations provide an approximately uniform full-sky coverage.

For model~3, we find a total of only $\sim3$ neutrinos above $10^{17}$~eV in 10 years from all 361 FRIIs of the \citet{Machalski2021} atlas combined, already indicating that the detection of individual sources is expected to be difficult. The largest contribution is expected from 3C284 ($z=0.23$), with 0.22 neutrinos, although the $L_{\rm kin}$ value estimated for this source by \citet{Machalski2021} is much larger than that of \citet{Godfrey2013} and appears to be an outlier (see Fig.~\ref{Lkin}). MRC1138-262 ($z = 2.1$), which has the largest $L_{\rm kin}$ value in the \citet{Machalski2021} atlas, would yield only $\sim 0.13$ events in 10 years. These examples confirm that individual FRII galaxies are generally expected to yield substantially fewer than one detected neutrino over 10 years.

Cygnus~A deserves a separate comment. Assuming $L_{\rm kin}=4.7\,10^{45}\,\rm erg\,s^{-1}$, it would yield $\sim 0.24$ neutrinos in 10 years. If instead one assumes $L_{\rm kin} = 2\,10^{46}\,\rm erg\,s^{-1}$ in the framework of model~3, the expected number increases to $\sim 3$ neutrinos in 10 years. However, as discussed in the previous subsection, the associated UHECR counterpart would then likely be challenged by the Auger composition constraints. Overall, obtaining on average more than one neutrino from a single source over 10 years would require an unrealistically nearby and/or exceptionally luminous FRII galaxy.

Although individual FRII galaxies are therefore unlikely to be detected as neutrino point sources, a statistical correlation with a population of bright radio galaxies may still be possible. To investigate this possibility, we concentrate on FRII galaxies with $L_{1.4}\geq10^{28}\,\rm W\,Hz^{-1}$. These galaxies combine two attractive properties: they contribute significantly to the expected diffuse neutrino flux (Fig.~\ref{Neut_contrib}) while having a very low space density (Fig.~\ref{Density}). Moreover, their high radio luminosities should make it possible in the future to establish relatively complete catalogs out to large redshifts, making them prime targets for correlation studies.

The left panel of Fig.~\ref{fraction_28} shows the average (over the 1000 realizations) fractional contribution of FRII galaxies with $L_{1.4}\geq10^{28}\,\rm W\,Hz^{-1}$ to the diffuse neutrino flux as a function of redshift for model~3, while the right panel shows the corresponding average number of sources within the enclosed volume. For instance, the $\sim 11$ FRII galaxies with $L_{1.4} \geq 10^{28}\,\rm W\,Hz^{-1}$ within $z<0.6$ contribute on average only one neutrino among the $\sim135$ expected to be detected by GRAND in 10 years. Extending the horizon to redshifts of 1, 2 and 5 increases this contribution to only $\sim2.5$, $\sim15$ and $\sim40$ neutrinos, shared among approximately $\sim60$, $\sim1750$ and $\sim10000$ FRII galaxies, respectively. These numbers suggest that identifying FRII galaxies as the sources of UHE neutrinos, for instance through a statistically significant correlation with source catalogs, will remain challenging. A more quantitative assessment would require both more complete FRII catalogs and more precise estimates of the expected angular correlation between cosmogenic neutrinos and their parent FRII galaxies, together with the angular resolution of GRAND or any other neutrino observatory, which is beyond the scope of this work. We also note that FRII galaxies with $10^{27}\leq L_{1.4}\leq10^{28}\,\rm W\,Hz^{-1}$, although contributing slightly more to the diffuse neutrino flux, are not expected to be better candidates because their space density is about one order of magnitude larger (Fig.~\ref{Density}).

\subsection{UHECR acceleration in relativistic jets}
\label{FRI}

An important question raised by the present work is whether our conclusions depend critically on the assumption that UHECR acceleration takes place in FRII lobes. An obvious alternative is that acceleration occurs instead in relativistic jets, in which case both FRI and FRII radio galaxies become potential contributors to the observed UHECR flux. Such scenarios have been investigated extensively in recent years (e.g., \citealt{Eichmann2018, Rodrigues2021, Eichmann2022, Seo2023, Seo2024, Seo2025}), generally with the aim of explaining the entire UHECR spectrum above the ankle.

One of the main motivations for these models is the proximity of the FRI radio galaxy Cen~A, in the direction of which Auger has reported a significant excess of UHECR events at intermediate angular scales \citep{AugerAniso2022}. Although this excess is naturally compatible with a contribution from Cen~A, it does not constitute compelling evidence for jet acceleration, as other interpretations remain possible (see, e.g., \citealt{Ding2021, Allard2024}). More generally, jet-dominated scenarios are subject to stronger constraints from UHECR anisotropy measurements than the FRII-lobe scenario considered in this work. Given the low local density of the relevant radio-galaxy population ($<10^{-4}\,\rm Mpc^{-3}$), the observed UHECR flux would be expected to be dominated by a handful of nearby sources, primarily Cen~A, M87 and Fornax~A, making the relatively weak anisotropies measured by Auger non-trivial to reproduce. Detailed studies, such as \citet{Eichmann2022}, have shown that agreement with the present data can nevertheless be achieved, but only for specific combinations of source properties (acceleration efficiency, activity history, etc.) and magnetic deflections during propagation. Such free parameters are physically well motivated and are in fact expected to vary significantly from one radio galaxy to another, as suggested by the substantial scatter around the $L_{1.4}$--$L_{\rm kin}$ relation shown in Fig.~\ref{Lkin}. It is nevertheless clear that the combinations of source parameters required to reproduce the present UHECR data also depend on other ingredients of the calculation, in particular the adopted Galactic magnetic-field model. Independent observational constraints on both the source properties and the Galactic magnetic field will therefore be essential before lending strong support to any particular realization of this class of models.

To assess the implications of this alternative scenario, we performed a simple exploratory calculation by extending the framework developed in this paper to all radio galaxies. The details are presented in Appendix~\ref{AllRG}. We find that, if FRI galaxies follow a different $L_{\rm kin}(L_{1.4})$ relation than FRII galaxies, as suggested by several studies of radio-galaxy jet energetics, relativistic jets can naturally account for a substantial fraction, or even the bulk, of the observed UHECR flux, in agreement with the conclusions of \citet{Rodrigues2021}. By contrast, the predicted cosmogenic neutrino flux remains remarkably similar to that obtained in our reference FRII-lobe scenario, reflecting the dominant contribution of the most luminous FRII galaxies to the diffuse neutrino background. These exploratory calculations therefore suggest that cosmogenic neutrinos alone are unlikely to distinguish between lobe- and jet-dominated acceleration scenarios. UHECR observations, on the other hand, retain a much stronger discriminating power, both because the predicted spectrum and composition differ and because anisotropy measurements provide important additional constraints on jet-dominated models.

\section{Summary and observational prospects}

In this work, we investigated the contribution of FRII radio galaxies to the ultra-high-energy cosmic-ray, cosmogenic neutrino and cosmogenic photon fluxes at Earth, assuming that UHECRs are accelerated in their radio lobes. Our model combines a recent determination of the luminosity-dependent density evolution of FRII galaxies, recent estimates of the relation between radio luminosity and jet kinetic power, and standard assumptions on the UHECR output of these sources.

Within this framework, we find that FRII galaxies can contribute at most about $10\%$ of the observed UHECR flux between $10^{18}$ and $10^{19}$~eV for the models considered here. The associated cosmogenic photon flux remains compatible with the Fermi-LAT constraints on the EGRB, confirming the conclusions of \citet{Globus2017}. Most importantly, the predicted cosmogenic neutrino flux is compatible, within $1\sigma$, with that inferred from the KM3-230213A event for models with $E_{\rm max}^{\rm ref}\ge10^{20}$~eV, making FRII radio galaxies a plausible origin for the first detected ultra-high-energy neutrino.

The predicted fluxes naturally depend on several assumptions of our model. The simplest dependence concerns $\epsilon_{\rm cr}$, the fraction of the jet kinetic luminosity transferred to cosmic rays (Sect.~2): all the predicted fluxes scale linearly with this parameter.

Concerning the UHECR spectrum injected by the sources, we adopted the standard assumption of non-relativistic diffusive shock acceleration with a spectral index $\beta=2$. The fraction of the kinetic energy transferred to UHECRs depends sensitively on this spectral index. However, the implications for the predicted UHECR, neutrino and photon fluxes at Earth are different depending on whether a deviation from the canonical $\beta = 2$ spectrum originates from the acceleration mechanism itself or from propagation and escape within the source environment. Appendix~\ref{HPF} illustrates this distinction by considering a simple energy-dependent escape model. Although such a high-pass-filter effect modifies the spectrum of escaping UHECRs, we find that its impact on the predicted neutrino flux around the energy of KM3-230213A remains moderate and becomes progressively weaker at higher energies, supporting the robustness of our main conclusions.

The observational consequences of this scenario can be grouped into three complementary categories.

First, future neutrino observatories such as GRAND 200k should collect enough cosmogenic neutrino events above $10^{17}$~eV to characterize the shape of the UHE neutrino spectrum. In our models, about $\sim50$ events are expected over ten years for model~2 and $\sim 135$ for model~3. However, as discussed in Sect.~\ref{Neutpoint}, detecting individual FRII sources appears out of reach, and establishing statistically significant correlations with bright FRII catalogs is likely to remain challenging.

The observational prospects are somewhat more favorable for UHECRs. Among all FRII galaxies, Cygnus~A stands out as the main realistic candidate for producing a detectable anisotropy signal. Unfortunately, it is also the most uncertain source in our sample, making quantitative predictions particularly difficult.

Two major sources of uncertainty dominate. First, the Galactic and extragalactic magnetic fields between Cygnus~A and the Earth remain poorly constrained. Although they are unlikely to isotropize the arrival directions or introduce sufficiently large delays to erase the proton signal above $3\,10^{19}$~eV, they still prevent reliable quantitative predictions of the expected anisotropy. Second, and probably more importantly, Cygnus~A is itself a rather peculiar FRII galaxy compared with more distant sources of similar radio luminosity \citep{Carilli1996}. As discussed in Sect.~\ref{UHECRpoint}, estimates of its kinetic luminosity span more than one order of magnitude, resulting in a correspondingly wide range of possible contributions to the observed UHECR flux and anisotropy.

More generally, the capability of Cygnus~A to accelerate particles up to the highest observed energies remains an open question. Since the pioneering work of \citet{Rachen1993}, a number of studies based on numerical simulations and multiwavelength observations have reached contrasting conclusions. For instance, \citet{Araudo2018} argued, mainly from the interpretation of radio observations of the primary hotspot, that Cygnus~A (and, by extension, FRII hotspots) is likely to be an inefficient UHECR accelerator, whereas the particle-in-cell simulations of \citet{Cerutti2023} lead to a more optimistic picture.

An important consequence of the dominant role played by Cygnus~A in the UHECR spectrum is that the predicted cosmogenic neutrino flux no longer uniquely determines the associated UHE proton abundance. As discussed in Sect.~\ref{UHECRpoint}, changing the value of $L_{\rm kin}$ assumed for Cygnus~A changes the relative contribution of protons above $3\,10^{19}$~eV from $0.8\%$ to about $60\%$, while leaving the predicted diffuse neutrino flux almost unchanged. The models discussed here can therefore accommodate a wide range of proton abundances above $10^{19}$~eV.

This behaviour differs significantly from the situation discussed by \citet{Decerprit2011}. In that work, for a given cosmological evolution of the sources and a given UHECR luminosity density, the cosmogenic neutrino flux was directly proportional to the proton contribution at a given energy. The origin of this difference is straightforward. Our scenario explicitly considers a population of rare discrete sources, taking into account both the luminosity function and the actual distribution of nearby objects. By contrast, \citet{Decerprit2011} assumed a continuous source distribution, which cannot describe the potentially dominant contribution of the nearest sources. As illustrated here by Cygnus~A (and more generally expected for any sufficiently rare source population\footnote{In the case of transient sources, the potential contribution of the nearest sources depends not only on the event rate but also on the magnetic-field-induced time delays of cosmic rays, making the notion of ``rare'' more subtle.}), nearby sources can substantially modify the observed UHECR spectrum while leaving the cosmogenic neutrino flux almost unchanged.

Future observations nevertheless offer several opportunities to test this scenario.

In the near future, the improved sensitivity to cosmic-ray composition provided by the Pierre Auger Observatory upgrade \citep{AugerUpgrade2020}, together with possible improvements in anisotropy analyses, should place stronger constraints on the UHE proton output of Cygnus~A and on the possible existence of an additional UHE proton component. Such a component has recently been advocated by several authors to improve the description of UHECR data. For instance, \citet{Muzio:2025gbr}, \citet{Alhebsi:2026bdk} (see also references therein) have explored the implications of the KM3NeT event detection on such a subdominant population of sources using HE neutrino, UHECR and diffuse gamma-ray background data by doing a global fit of the parameters defining generic main and secondary sources populations (but not describing actual known astrophysical object populations). Although the most recent Auger composition analyses (e.g., \citet{Tkachenko2025}) clearly allow for a non-zero proton abundance above $10^{19}$~eV, current constraints on the existence of a distinct UHE proton component remain rather weak\footnote{To the best of our knowledge, no upper limit on the proton relative abundance in the highest-energy bins has been published so far.}. Establishing such a component would not by itself validate our scenario or imply a connection with KM3-230213A, but it would provide an important constraint on models of this kind, just as more stringent upper limits would.

Over longer timescales, direct observations of radio galaxies themselves may provide even more decisive tests. Multiwavelength studies of jets and lobes with new facilities such as SKA (\citet{Hardcastle2026} and references therein) should considerably improve our understanding of jet composition and particle acceleration in these systems. Such observations may ultimately provide the most direct way to test the physical assumptions underlying the scenario proposed here, as well as many related models.


For all these reasons, although the scenario discussed in this paper provides a possible explanation for the origin of the KM3-230213A neutrino event, it will probably remain difficult to test or rule out conclusively in the next few years using UHECR or neutrino observations alone. Ultimately, the combination of multimessenger observations with improved astrophysical constraints on particle acceleration and jet composition in powerful radio galaxies should provide a coherent framework within which scenarios such as the one proposed here can be critically tested.

\begin{acknowledgements}
      D.A. wishes to thank G.~Decerprit and N.~G. Busca for their participation to previous related works. B.B. wishes to thank A. Condorelli for fruitful discussions. N.G. gratefully acknowledges the support of the Simons Foundation (MP-SCMPS-00001470, 
N.G.). DA, BB and EP acknowledge financial support from the Centre national d’études spatiales (CNES), France (ROR: https://ror.org/04h1h0y33) within the framework of the EUSO mission and from the INTERCOS master project of IN2P3.
\end{acknowledgements}

%


\begin{appendix} 
\section{Impact of cosmic magnetic fields on UHECRs from Cygnus~A}
\label{Deflections}

\begin{figure*}[ht!]
        \centering

        \includegraphics[width=8cm]
        {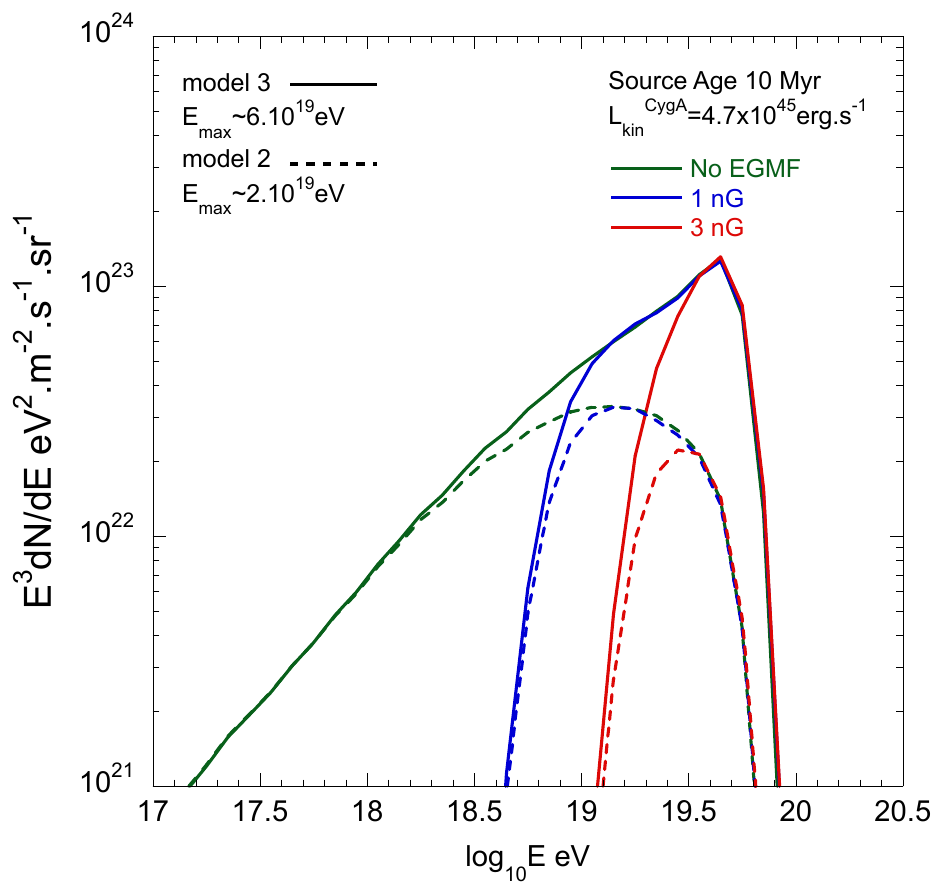}
        \includegraphics[width=8cm]
        {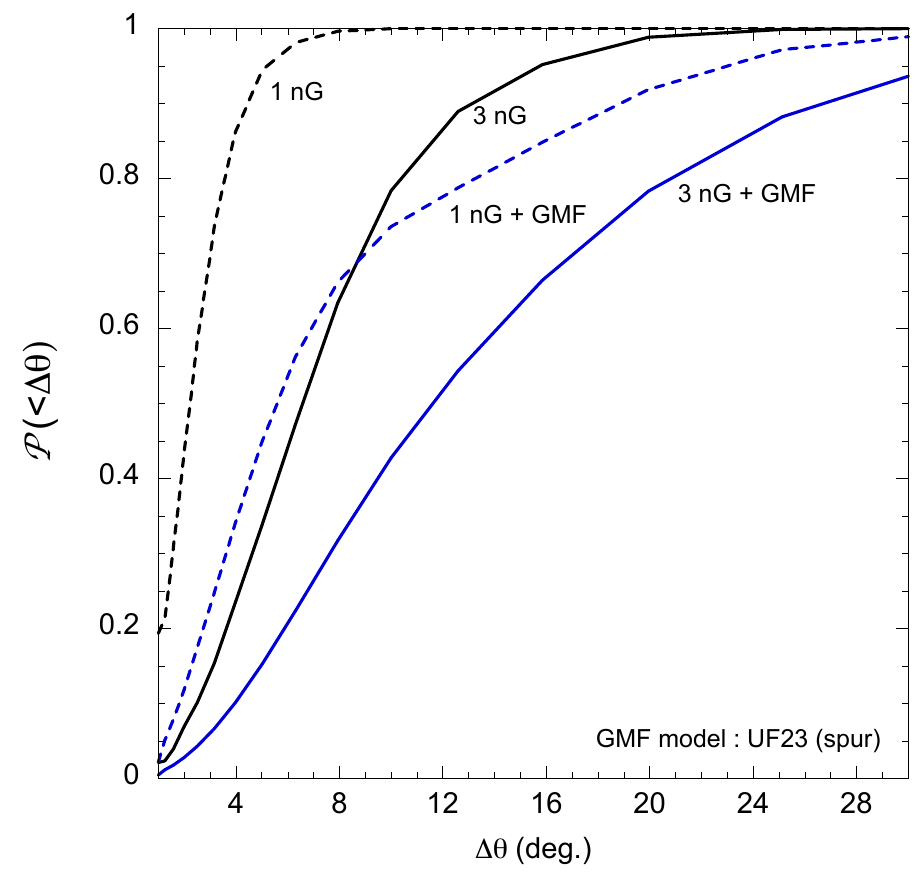}
        
        \caption{Left : Spectrum at the Earth of UHECRs emitted by Cygnus A for models 2 and 3 for 3 different assumptions for the EGMF value, no EGMF, 1 nG and 3 nG (see legend). A 10 Myr, source age is assumed. Right : Cumulative function of the angular Deflections suffered by UHECR arriving at the earth from Cygnus A with $E>32$~EeV. Different EGMF values are considered with and without the contribution of the Galactic magnetic field (see labels).}
        \label{CygA_timeDelays}
    \end{figure*}

In Sect.~\ref{UHECRpoint}, we summarized the main conclusions regarding the impact of cosmic magnetic fields on the observability of UHECRs from Cygnus~A. In this Appendix, we present the calculations supporting these conclusions.

We consider the reference value $L_{\rm kin}=4.7\,10^{45}\,\rm erg\,s^{-1}$ adopted throughout the paper. In the calculations of Sect.~\ref{Sect:Diffuse}, we assumed a negligible EGMF. While this assumption has little impact on the diffuse UHECR and cosmogenic neutrino fluxes, it becomes much more relevant when discussing the observability of individual sources. Cosmic magnetic fields affect both the angular extension of the UHECR image and the time delays with respect to the electromagnetic emission.

FRII radio lobes are transient structures, with typical ages ranging from a few $10^6$ to about $10^8$~yr (see e.g.~\citet{Machalski2021, Rafferty2006, Birzan2008, Ito2008}). Magnetic time delays may therefore reduce the observed UHECR flux from an individual source whenever they become comparable to the source lifetime.

To quantify these effects, we computed the propagation of UHECRs emitted by Cygnus~A assuming the Model~3 scaling for $E_{\rm max}$ (yielding $E_{\rm max}^{\rm Cygnus\,A}\simeq6\times10^{19}$~eV), different EGMF strengths (1 and 3~nG), and various source ages, using the numerical code described in \citet{Globus2008}. The EGMF is assumed to be a purely turbulent field with a maximum turbulence scale of 1~Mpc.

Figure~\ref{CygA_timeDelays} summarizes the two main consequences of the EGMF. The left panel illustrates the impact of magnetic time delays on the UHECR spectrum reaching the Earth, while the right panel shows the expected cumulative distribution of angular deflections above 32~EeV.

The results are shown in the left panel of Fig.~\ref{CygA_timeDelays}, assuming a source age of 10~Myr, which is shorter than most of the age estimates reported in the above references. As expected, the impact of magnetic time delays increases with the strength of the EGMF. Nevertheless, the predicted UHECR spectrum at the Earth remains almost unchanged with respect to the case without EGMF between $\sim2\times10^{19}$ and $\sim5\times10^{19}$~eV, where the spectrum is expected to terminate because of the GZK suppression, even for an EGMF as large as 3~nG.

Since Cygnus~A is located at the center of a rich galaxy cluster, the magnetic field in its immediate environment may well be much stronger than a few nG, possibly reaching several hundred nG over scales of a few hundred kpc. While such localized magnetic fields may induce substantial time delays for low-rigidity cosmic rays, they are not expected to affect significantly UHECR protons above $10^{19}$~eV, as shown for example by \citet{Kotera2009} for transient AGN jets embedded in galaxy clusters.

The right panel of Fig.~\ref{CygA_timeDelays} shows the cumulative distribution of angular deflections expected for protons above 32~EeV arriving from Cygnus~A. Considering only the EGMF, the deflections remain rather small: even for a 3~nG field, about $80\%$ of the protons are deflected by less than $10^\circ$. Including the Galactic magnetic field (GMF) increases the deflections, although they remain modest compared to those expected for heavier nuclei at the same energy. The examples shown in Fig.~\ref{CygA_timeDelays} adopt one of the recent GMF models proposed by \citet{Unger2024} (hereafter UF23). In the 3~nG case, about $50\%$ of the protons are deflected by less than $12^\circ$, while nearly $90\%$ remain within $30^\circ$ of the source direction.

Moreover, for directions close to Cygnus~A, most recent GMF models, including \citet{JF2012}, UF23 and \citet{Xu2024}, predict a magnetic magnification\footnote{i.e., an amplification of the flux due to magnetic lensing (see e.g.~\citet{Harari2002}).} by a factor of roughly 2--5 for rigidities around $3\times10^{19}$~V. The main exception is the model proposed by \citet{KTS2024}, which predicts a magnification closer to unity in this region of the sky.

Overall, the calculations presented in this Appendix support the conclusions summarized in Sect.~\ref{UHECRpoint}: neither magnetic time delays nor magnetic deflections appear to be sufficient to qualitatively modify the expected observability of UHECRs from Cygnus~A above a few $10^{19}$~eV.

\section{impact of the shape of the UHECR output from FRII galaxies}
\label{HPF}

The calculations presented in the main text assume the canonical $\beta=2$ power-law expected from test-particle diffusive shock acceleration. In practice, however, the spectrum of UHECRs escaping from FRII galaxies may differ from this reference case for different physical reasons. It may reflect a genuinely different acceleration spectrum, or result from energy-dependent escape from the source environment.

These two situations have different implications for the expected UHECR, neutrino and photon fluxes. In this appendix, we briefly illustrate these differences by considering both a change of the accelerated spectrum and a simple high-pass-filter model for the escape process.

Changing the spectral index of the accelerated cosmic rays modifies the fraction of the energy budget transferred to UHECRs (cosmic rays above $10^{17}$~eV), and therefore the expected UHECR, neutrino and photon fluxes at Earth. In the main text, we adopted the canonical value $\beta=2$. The corresponding dependence of the UHECR energy fraction on $\beta$ is given by Eq.~\ref{Eq:omega_cr}. Assuming $E_{\rm max}=10^{20}$~eV, spectral indices of 1.8 and 2.1 would correspond to values of $\omega$ approximately three times larger and three times smaller, respectively, than for the reference case $\beta=2$, leading to similar changes in the predicted fluxes.

A qualitatively different situation arises if the escaping UHECR spectrum is modified by energy-dependent escape from the source environment. For instance, magnetic confinement in the acceleration region may act as a high-pass filter, since higher-energy particles escape more efficiently than lower-energy ones. Such an effect hardens the spectrum of escaping UHECRs without modifying the acceleration process itself.

Unlike a genuine change of the acceleration spectrum, such a high-pass-filter effect does not modify the fraction of the source power transferred to UHECRs. Instead, it changes the relative contributions of escaping and confined cosmic rays to the production of secondary particles. The resulting neutrino and photon fluxes therefore depend on the fate of the trapped UHECRs and on the way they lose their energy within the source environment.

To illustrate this effect, we implemented modified versions of models~2 and~3 by introducing an energy-dependent escape probability below $E_{\rm max}$. This hardens the escaping spectrum to effective indices $\beta_{\rm esc} = 1$ and $\beta_{\rm esc} = 0$, which are compared with the reference case $\beta_{\rm esc}=\beta=2$ adopted in the main text.

Figure~\ref{HPF} shows the resulting cosmogenic neutrino fluxes, considering only the contribution of escaping UHECRs. The impact of the high-pass-filter effect is moderate around the energy of the KM3-230213A event and becomes progressively weaker at higher energies. This simply reflects the correlation between neutrino and parent-UHECR energies: the highest-energy neutrinos originate from the highest-energy cosmic rays, which are the least affected by the high-pass filter.


The diffuse neutrino fluxes calculated here include only the contribution of escaping UHECRs. They may therefore represent lower limits, since UHECRs confined within the source could also produce neutrinos through interactions either with the CMB or with ambient photon fields, provided they interact before losing their energy through adiabatic expansion in the source environment.\footnote{Addressing this question would require a detailed model of particle acceleration, energy losses and escape in FRII hotspots, which lies beyond the scope of the present work.}

The high-pass-filter effect would also modify the predicted UHECR spectrum, although mainly below $10^{19}$~eV, and would therefore not qualitatively affect our discussion of the contribution of FRII galaxies above $3\,10^{19}$~eV. Cosmogenic photons are expected to be more strongly affected, since they are predominantly produced through pair production by lower-energy UHECRs, for which the high-pass-filter effect is strongest.

\begin{figure}[ht!]
        \centering
        \includegraphics[width=8cm]
        {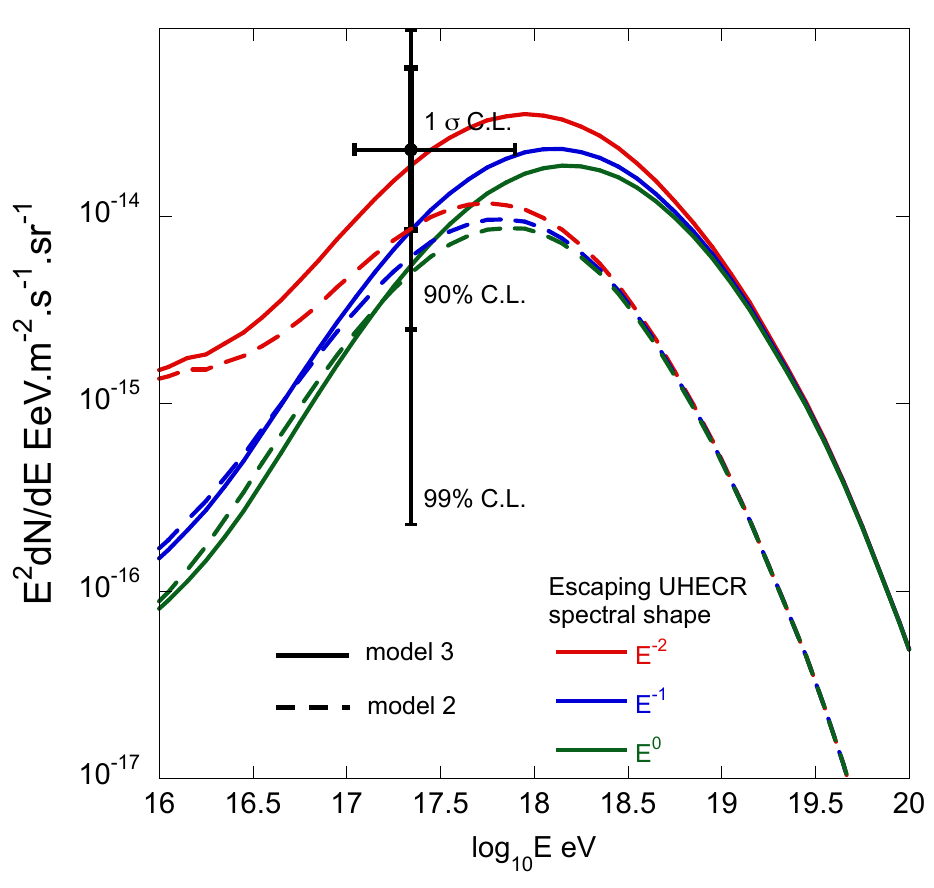}
        \caption{Impact of a possible high-pass-filter effect, due to the escape process from the source environment, on the expected cosmogenic neutrino flux in the cases of model 2 (dased lines) and 3 (lines). In addition to the unmodified case, $\beta_{\rm esc}=\beta=2$, the cases $\beta_{\rm esc}=1$ and $\beta_{\rm esc}=0$ are shown (see legend).}
        \label{HPF}
    \end{figure}

\section{UHECR and cosmogenic neutrino fluxes from all radio galaxies (relativistic jets scenario)}
\label{AllRG}

\begin{figure*}[ht!]
        \centering

        \includegraphics[width=8cm]
        {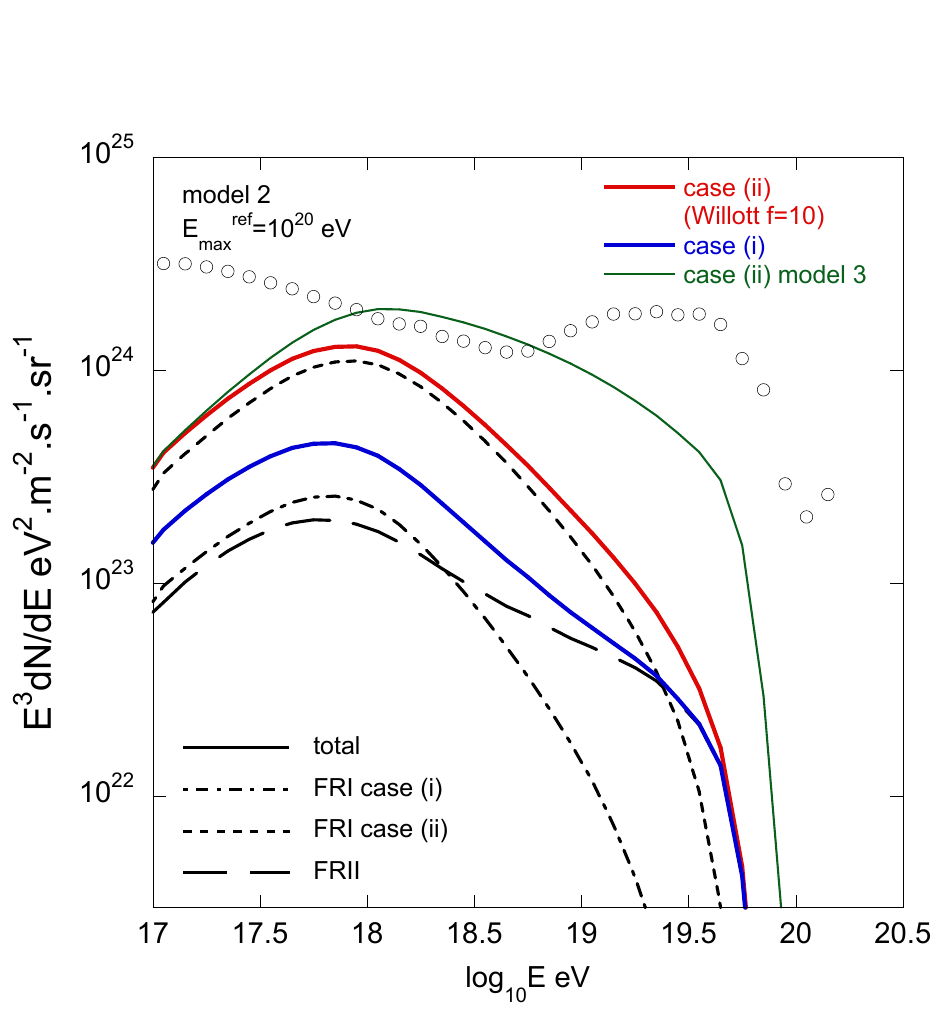}
        \includegraphics[width=8cm]
        {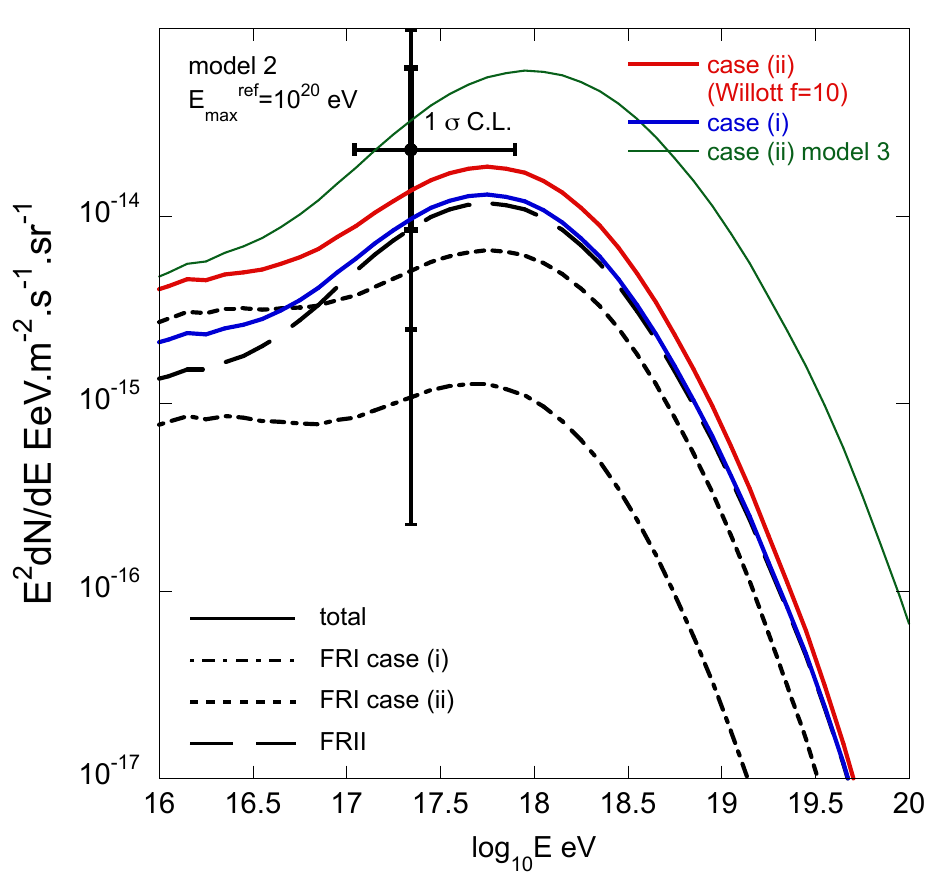}
        
        \caption{Left: Predicted diffuse UHECR spectrum from all radio galaxies assuming that particle acceleration takes place in relativistic jets. The contributions of FRII galaxies and FRI galaxies for cases (i) and (ii), together with their sum, are shown for model~2 (see legend). For model~3, only the total flux for case (ii) is displayed. Right: Corresponding diffuse cosmogenic neutrino fluxes for the same cases.}
        
        \label{FRI_FRII}
    \end{figure*}

Throughout this paper, we have focused on a scenario in which UHECRs are accelerated in the lobes of FRII radio galaxies. A natural alternative is that particle acceleration instead takes place in relativistic jets. In this case, both FRI and FRII radio galaxies become potential contributors to the UHECR population. Although a detailed investigation of this scenario lies beyond the scope of the present work, it is nevertheless instructive to estimate its implications for the diffuse UHECR and cosmogenic neutrino fluxes, and to compare them with those obtained in our reference lobe scenario.

To this end, we construct a simple model based on the same ingredients as those adopted throughout this paper. We keep the same source spectrum, the same reference maximum energy $E_{\rm max}^{\rm ref}$ defined at $L_{\rm kin}=10^{47}\,\rm erg\,s^{-1}$, and the same assumptions regarding the UHECR output. The only modification concerns the relation between the kinetic power and the radio luminosity of FRI galaxies. We consider two possibilities:

(i) FRI galaxies follow the same $L_{\rm kin}(L_{1.4})$ relation as FRII galaxies, given by Eq.~\ref{Eq:Lkin}, which was fitted to the estimates of \citet{Machalski2021};

(ii) FRI galaxies instead follow the \citet{Willott1999} relation with $f=10$.

The second hypothesis is motivated by several studies of radio-galaxy jet energetics. As discussed by \citet{Smolcic2017}, low-luminosity radio galaxies, which are predominantly FRI objects, tend to favour values of $f\simeq 10$--15 when interpreted within the \citet{Willott1999} framework, whereas the estimates of \citet{Daly2012} and \citet{Machalski2021} for FRII galaxies are incompatible with such large values. Assuming $f=10$ therefore implies kinetic powers approximately three times larger than those inferred from the \citet{Machalski2021} relation (and about four times larger than for $f=4$). Within our model, this translates into UHECR luminosities roughly three times larger and maximum energies higher by a factor $\sqrt{3}$ for a given radio luminosity. Such a difference could naturally arise from a larger hadronic content of FRI jets, as discussed in Sect.~\ref{UHECRoutput}.

Since we keep all the other hypotheses of Sect.~\ref{ModelHypotheses} identical to those adopted for the FRII lobe scenario, the different $L_{\rm kin}(L_{1.4})$ relation assumed for FRI galaxies implies, for a given radio luminosity, a UHECR luminosity approximately three times larger and a maximum energy higher by a factor $\sqrt{3}$.

Before discussing the results, it is worth emphasizing that the present calculations are only intended to assess the overall energetics of this alternative scenario rather than to reproduce the observed UHECR spectrum in detail. A realistic model describing the entire UHECR spectrum above the ankle should also include the nuclei, which dominate the highest-energy cosmic rays according to Auger composition analyses. Here, we consider only protons since our purpose is simply to determine whether this alternative scenario involving all radio galaxies could account for the observed UHECR energetics\footnote{In other words, to determine whether this scenario involving all radio galaxies could reproduce the overall UHECR flux, contrary to the reference model involving only FRII lobes.}. Moreover, besides the cosmological evolution of the sources, the cosmogenic neutrino flux depends primarily on the maximum energy per nucleon. Restricting the calculations to the proton component is therefore sufficient for the present discussion. We also note that, as for most candidate UHECR sources, there is currently no robust prediction regarding the composition of particles accelerated in radio-galaxy jets, nor whether such models would naturally reproduce the composition inferred by Auger.

We keep the same procedure as for FRII galaxies, randomly sampling the LDDE presented in Sect.~\ref{Sect:LDDE} while accounting for the luminosity-dependent fraction of FRI galaxies among the radio-galaxy population. Nearby FRI galaxies from the catalog of \citet{vanVelzen2012}, in particular Cen~A, M87 and Fornax~A, are included in every realization, using their measured radio luminosities or kinetic-power estimates whenever available. The results are presented in Fig.~\ref{FRI_FRII}.

The predicted UHECR fluxes are shown in the left panel of Fig.~\ref{FRI_FRII}. The calculations are presented for model~2 ($E_{\rm max}^{\rm ref}=10^{20}$~eV), together with model~3 ($E_{\rm max}^{\rm ref}=3\,10^{20}$~eV) for case~(ii). Overall, these calculations indicate that a scenario in which UHECRs are accelerated in relativistic jets of all radio galaxies can naturally account for a substantial fraction, or even the bulk, of the observed UHECR flux.

The two assumptions adopted for FRI galaxies nevertheless lead to markedly different predictions. In case~(i), where FRI galaxies follow the same $L_{\rm kin}(L_{1.4})$ relation as FRII galaxies, the predicted total flux remains well below the spectrum measured by Auger, reaching at most about $25\%$ of the observed flux around $10^{18}$~eV. At these energies, FRI galaxies contribute slightly more than FRII galaxies because radio galaxies with $L_{1.4}<10^{26}\,\rm W\,Hz^{-1}$ dominate the local luminosity density in the adopted LDDE. The FRI contribution, however, cuts off at lower energies because these lower-luminosity objects also reach lower maximum energies.

By contrast, case~(ii), based on the \citet{Willott1999} relation with $f=10$, predicts a substantially larger FRI contribution. The corresponding UHECR flux is approximately three times higher than in case~(i), extends to higher energies, and becomes much more compatible with the Auger measurements, although it still naturally softens above $10^{18}$~eV owing to the luminosity function of FRI galaxies. Model~3 slightly overshoots the observed spectrum because of the larger source maximum energies.

These simple estimates therefore suggest that relativistic jets of radio galaxies constitute a viable candidate for producing most of the observed UHECR flux, provided that nuclei are accelerated together with protons and escape from the source environment with maximum energies scaling approximately as the nuclear charge $Z$. Although reproducing the detailed UHECR spectrum and composition would require a more refined model, the natural softening of the proton component above a few $10^{18}$~eV is qualitatively consistent with the composition evolution inferred by Auger. This conclusion agrees with previous studies \citep{Eichmann2018, Rodrigues2021, Eichmann2022, Seo2025}. In particular, the need for larger kinetic powers in FRI galaxies than in FRII galaxies, for a given radio luminosity, is fully consistent with the results of \citet{Rodrigues2021}.

The corresponding cosmogenic neutrino fluxes are shown in the right panel of Fig.~\ref{FRI_FRII}. In contrast to the UHECR predictions, the neutrino fluxes obtained for cases~(i) and (ii) are remarkably similar. This is because the diffuse cosmogenic neutrino production remains dominated by the most luminous FRII galaxies, whose higher maximum energies and stronger cosmological evolution largely compensate for the much larger number of lower-luminosity FRI galaxies.

In case~(ii), the contribution of FRI galaxies to the neutrino flux becomes non-negligible, since the larger kinetic powers assumed for these sources imply higher maximum energies for a given radio luminosity. Nevertheless, the total cosmogenic neutrino flux remains very close to that predicted in our reference lobe scenario, which is expected because the dominant contribution still originates from the FRII population.

Overall, these exploratory calculations suggest that moving the assumed acceleration site from FRII lobes to relativistic jets of all radio galaxies has a strong impact on the predicted UHECR component while leaving the cosmogenic neutrino flux comparatively unchanged. Cosmogenic neutrinos alone are therefore unlikely to discriminate between these two scenarios. By contrast, UHECR observations---through their spectrum, composition and anisotropies---retain a much stronger discriminating power and would provide the most direct observational test of such a jet-dominated scenario.

\end{appendix}

\end{document}